# Penalty Dual Decomposition Method For Nonsmooth Nonconvex Optimization—Part I: Algorithms and Convergence Analysis

Qingjiang Shi and Mingyi Hong

*Abstract*—Many contemporary signal processing, machine learning and wireless communication applications can be formulated as nonconvex nonsmooth optimization problems. Often there is a lack of efficient algorithms for these problems, especially when the optimization variables are nonlinearly coupled in some nonconvex constraints. In this work, we propose an algorithm named penalty dual decomposition (PDD) for these difficult problems and discuss its various applications.

The PDD is a double-loop iterative algorithm. Its inner iterations is used to inexactly solve a nonconvex nonsmooth augmented Lagrangian problem via block-coordinate-descent-type methods, while its outer iteration updates the dual variables and/or a penalty parameter. In Part I of this work, we describe the PDD algorithm and rigorously establish its convergence to KKT solutions. In Part II we evaluate the performance of PDD by customizing it to three applications arising from signal processing and wireless communications.

## I. INTRODUCTION

Many important engineering problems arising from signal processing, wireless communications and machine learning can be modeled as nonconvex nonsmooth optimization problems. These problems are generally difficult to solve especially when the optimization variables are nonlinearly coupled in some (possibly nonconvex) constraints. This two-part paper provides a algorithmic framework that can fully exploit the problem structure, for optimizing a nonconvex nonsmooth function subject to nonconvex but continuously differentiable *coupling* constraints.

Nonconvex problems with constraints that couple a few design variables often arise in contemporary applications. For example, in the joint source-relay design of many multiple-input-multiple-output (MIMO) relay systems [2]–[4], the relay power constraints often couple the source or relay precoders in a *bi-quadratic* manner, meaning that fixing one variable (i.e., the source precoders), then the constraint function becomes qudratic with respect to the other variable (i.e., the relay recoders). Another popular example arise in the family of quality-of-service (QoS)-constrained power minimization problems, in which the signal-to-interference-plus-noise ratio (SINR) functions or the (weighted) mean-square-error (MSE) functions are also quadratic in the beamformers [5]–[9]. In problems such as dictionary learning [10], [11], nonnegative matrix factorization [12]–[14], and geometry-based blind source separation [15], the variables are coupled in a *bi-linear* manner by certain equality constraints. Other problems with nonlinear and nonconvex constraints coupling can be found in [16]–[19]. Such constraint coupling makes developing efficient low-complexity, and parallel algorithms a very challenging task.

Generally speaking, when designing algorithms for an engineering problem, it is important to exploit, as much as possible, its fundamental structures in order to improve solution quality and/or speed. For problems with multi-blocks and coupling constraint, it is the *block structure* that often gets exploited. One such popular method is the alternating optimization (AO) method, which replaces difficult joint optimization over all variables with a sequence of easier optimization over individual (block) variable. For instance, for two-hop relay broadcast channel, the authors of [9] considered joint source-relay design for achieving power minimization subject to SINR constraints, where the source precoder and relay precoder are coupled with each other. Observing that the power minimization problem is convex with respect to the source precoder or the relay precoder, the work [9] used the AO method to address the power minimization problem. Similar to [9], the work [2] also used the AO method to address the joint source-relay design to achieve sum rate maximization in a MIMO relay interference channel. However, the AO method can only provide feasible solutions in the coupling constraint case and cannot guarantee convergence to stationary solutions (or KKT points) unless the objective has some special structure; see for example [8]. In particular, the AO method easily gets trapped in some unexpected points in the equality coupling constraint case; see [20] for illustrative examples. To deal with a special class of equality coupling constraint $\mathbf{Z} = \mathbf{XY}$ (where $\mathbf{X}$, $\mathbf{Y}$ and $\mathbf{Z}$ are all matrix variables) that arises from relay network design, the work [21] first transformed the equality coupling constraint into two matrix inequalities and then used concave-convex procedure to solve the resultant problem. However, this method is not only computationally expensive, but also lacks convergence guarantee to stationary solutions.

Another popular approach that can deal with the coupled constraint, especially the equality coupling constraints, is the penalty method [22]. The basic idea of penalty method is to move the difficult constraints to the objective function as a penalty term, so that infeasible points can get relatively high cost compared with the feasible ones. For example, in [17], Kuang *et al* used penalty method to approximate the solution of the symmetric nonnegative matrix factorization problem. In [3], Shi *et al* used penalty method to solve the joint source-relay design problem for full-duplex MIMO relay systems. The work [23] showed that penalty method can be applied to solve the rank minimization problem, an important class of problems that often arises from signal processing. However, penalty methods could be very inefficient, because it usually requires that certain penalty parameter goes to infinity, resulting in ill-conditioning for its subproblems. Augmented Lagrangian (AL) methods [24], [25] was proposed to overcome the limitations of penalty methods by introducing an additional dual-related term. In the AL methods, a sequence of *AL subproblems* (i.e., the problems of minimization of the augmented Lagrangian) needs to be exactly or approximately solved [22]. When the AL subproblems are easily solvable, the AL methods are attractive as they can be often easily implemented (often in a matrix-free manner) [26] and have at least local convergence guarantees under relatively mild assumptions [27], [28]. However, the AL subproblems are generally hard to solve especially when they have complicated constraints. Further, the AL method generally cannot deal with nonsmooth penalty function in the objective.

As an important variant of augmented Lagrangian method, alternating direction method of multipliers (ADMM) has recently regained popularity due to its applicability in many large-scale problems [29]. Differently from the standard AL method, a single iteration of *block coordinate descent* (BCD) or AO is used to *approximately* minimize the augmented Lagrangian at each iteration of ADMM. That is, the *AL subproblem* is minimized only *approximately*, by solving a sequence of smaller, and potentially easier, subproblems generated by the block coordinate decomposition. Indeed, it is the idea of combining block decomposition and approximate AL subproblem minimization that enables the ADMM to fully exploit the block structure of the problem. Although the ADMM has been widely used in the areas of signal processing [15], [30], [31], wireless communication [5], [7], [32], [33], and machine learning [14], [29], [34], [35], they are primarily developed for convex problems with linearly coupling constraints. Generally speaking, ADMM does not converge for nonconvex problems, except for a few special cases; see recent developments in [36]–[39] and the references therein.

In this work, we propose an optimization framework named penalty dual decomposition (PDD), which integrates the penalty mehtod, the AL method and the ADMM method. Specifically, our framework is a double-loop algorithm where the inner loop *approximately* solves the AL subproblem, while the outer loop updates the dual variable and/or a certain penalty parameter. To exploit the problem structure as fully as possible, a block-coordinate-descent (BCD) based method is used to approximately solve the AL subproblem. In Part I of the paper, we first introduce the notion of generalized gradient [40] and provide conditions under which a KKT point exists. We then rigorously prove the convergence of the PDD to KKT points under some constraint qualification (CQ) condition. Furthermore, to address AL problems with nonconvex constraints using BCD-type algorithms, we propose stochastic BSUM algorithm and prove its convergence. Our proof is critically dependent on the randomization introduced to the original BSUM algorithm, which provides the algorithm with good convergence behavior even in the presence of *nonconvex constraints*. In the second part of this paper, we customize the PDD to several engineering problems arising from signal processing and wireless communications. Our numerical results show that PDD outperforms a number of state-of-the-art algorithms, therefore validating the effectiveness of the PDD method in solving nonconvex nonsmooth problem with coupling constraints.

*Notations*: Throughout this paper, we use uppercase bold letters for matrices, lowercase bold letters for column vectors, and regular letters for scalars (unless otherwise specified). The notations $\mathbb{R}^n$, $\mathbb{R}^n_+$ and $\mathbb{R}^n_-$ denote the $n$-dimensional space of real number, nonnegative real number, nonpositive real number, respectively. For a vector $\boldsymbol{x}$, $\|\boldsymbol{x}\|$ and $\|\boldsymbol{x}\|_\infty$ denote Euclidean norm and element-wise infinity norm, respectively. $B_\delta(\boldsymbol{x}_0)$ denotes a Euclidean ball centered at $\boldsymbol{x}_0$ with radius $\delta$. For a scalar function $f(\cdot)$, $f'(\cdot)$ and $\nabla f(\cdot)$ respectively denote its derivative and gradient with respect to its argument. For a multivariate function $f(\boldsymbol{x}, \boldsymbol{y})$, $\nabla_{\boldsymbol{x}} f(\boldsymbol{x}, \boldsymbol{y})$ denotes its gradient with respect to $\boldsymbol{x}$. For vector functions $\boldsymbol{g}(\boldsymbol{x})$ and $\boldsymbol{h}(\boldsymbol{x}, \boldsymbol{y})$, $\nabla \boldsymbol{g}(\boldsymbol{x})$ denotes the Jacobian matrix of $\boldsymbol{g}(\boldsymbol{x})$ and $\nabla_{\boldsymbol{x}} \boldsymbol{h}(\boldsymbol{x}, \boldsymbol{y})$ denotes the Jacobian matrix of $\boldsymbol{h}(\boldsymbol{x}, \boldsymbol{y})$ with respect to $\boldsymbol{x}$. For a convex function $\hbar(\boldsymbol{x})$, $\partial \hbar(\boldsymbol{x})$ denotes its subdifferential. $T_{\mathcal{Z}}(\boldsymbol{z})$ and $N_{\mathcal{Z}}(\boldsymbol{z})$ denotes the tangent cone and normal cone [41] of the set $\mathcal{Z}$ at point $\boldsymbol{z}$, respectively, and these definitions are formally given in Appendix A. The notation $int\mathcal{Z}$ denotes the interior of the set $\mathcal{Z}$ while the notation $(\boldsymbol{x}_i)_i$ denotes a vector stacked by all subvectors $\boldsymbol{x}_i$'s.

## II. NONCONVEX NONSMOOTH OPTIMIZATION AND KKT CHARACTERIZATION

Consider the following problem

$$\begin{aligned} \min_{\boldsymbol{x} \in \mathcal{X}, \boldsymbol{y}} \quad & F(\boldsymbol{x}, \boldsymbol{y}) \triangleq f(\boldsymbol{x}, \boldsymbol{y}) + \sum_{j=1}^{n_y} \tilde{\phi}(\boldsymbol{y}_j) \\ \text{s.t.} \quad & \boldsymbol{h}(\boldsymbol{x}, \boldsymbol{y}) = \boldsymbol{0}, \\ & \boldsymbol{g}_i(\boldsymbol{x}_i) \leq 0, \; i = 1, 2, \ldots, n \end{aligned} \quad \text{(P)}$$

where

- the feasible set $\mathcal{X}$ is the Cartesian product of $n$ *closed convex* sets: $\mathcal{X} \triangleq \Pi_{i=1}^n \mathcal{X}_i$ with $\mathcal{X}_i \subseteq \mathbb{R}^{n_i}$ and $\sum_{i=1}^n n_i = N$;
- the optimization variable $\boldsymbol{x} \in \mathbb{R}^N$ is decomposed as $\boldsymbol{x} = (\boldsymbol{x}_1, \boldsymbol{x}_2, \ldots, \boldsymbol{x}_n)$ with $\boldsymbol{x}_i \in \mathcal{X}_i$ $i = 1, 2, \ldots, n$, and $\boldsymbol{y} \in \mathbb{R}^M$ is decomposed as $\boldsymbol{y}_j \in \mathbb{R}^{m_j}$, $j = 1, 2, \ldots, m$, with $\sum_{j=1}^m m_j = M$;
- $f(\boldsymbol{x}, \boldsymbol{y})$ is a scalar *continuously differentiable* function; $\tilde{\phi}(\boldsymbol{y}_j)$ is a composite function in the form of $\phi_j(s_j(\boldsymbol{y}_j))$, with each $s_j(\boldsymbol{y}_j)$ being a *convex* but possibly nondifferentiable function while $\phi_j(x)$ being a *nondecreasing* and *continuously differentiable* function;

- for each $i$, $\boldsymbol{g}_i(\boldsymbol{x}_i) \in \mathbb{R}^{q_i}$ is a vector of $q_i$ *continuously differentiable* functions, and we define $q \triangleq \sum_{i=1}^{n} q_i$;
- $\boldsymbol{h}(\boldsymbol{x}, \boldsymbol{y}) \in \mathbb{R}^p$ is a vector of $p$ *continuously differentiable* functions.
- The feasible set of problem $(P)$, given below, is nonempty

$$\mathcal{Z} \triangleq \{(\boldsymbol{x}, \boldsymbol{y}) \in \mathbb{R}^N \times \mathbb{R}^M \mid \boldsymbol{x} \in \mathcal{X}, \\ \boldsymbol{h}(\boldsymbol{x}, \boldsymbol{y}) = \boldsymbol{0}, \ \boldsymbol{g}_i(\boldsymbol{x}_i) \leq \boldsymbol{0}, \ \forall \ i\}. \quad (1)$$

In the above problem, the constraint coupling is mainly represented by the equality constraint $\boldsymbol{h}(\boldsymbol{x}, \boldsymbol{y}) = \boldsymbol{0}$, while for each $i$, the inequality constraint $\boldsymbol{g}_i(\boldsymbol{x}_i) \leq \boldsymbol{0}$ represents the possibly nonconvex constraints for $\boldsymbol{x}_i$. Note that we do not explicitly write down the constraint set for the block $\boldsymbol{y}$. This is because we assume that $\boldsymbol{y}$ only has convex constraints, and such convex constraints can be absorbed into the nonsmooth part of the objective $\tilde{\phi}(\boldsymbol{y}_j)$, by introducing indicator functions of convex sets.

Further, we remark that the term $\sum_{j=1}^{n_y} \tilde{\phi}(\boldsymbol{y}_j)$ represents the nonsmooth part of the objective function. Typically, the composite function $\tilde{\phi}(\boldsymbol{y}_j) = \phi_j(s_j(\boldsymbol{y}_j))$ can take the form of sparsity promoting functions. For instance, in the case of log-based sparsity promotion function, we have $\phi_j(z) = \lambda \log\left(1 + \frac{z}{\epsilon}\right)$ and $s_j(\boldsymbol{y}_j) = \|\boldsymbol{y}_j\|$, and thus $\tilde{\phi}(\boldsymbol{y}_j) = \lambda \log\left(1 + \frac{\|\boldsymbol{y}_j\|}{\epsilon}\right)$. Here $\lambda$ and $\epsilon$ are two positive sparsity-related control parameters. We refer readers to [42, TABLE I] for more examples of sparsity promotion functions, e.g, lasso penalty function, SCAD penalty function, etc. Since the term $\sum_{j=1}^{n_y} \tilde{\phi}(\boldsymbol{y}_j)$ could be neither convex nor differentiable, we need to use generalized gradient [43] to characterize the first-order optimality condition, which is the main topic of the following two sections.

*A. Preliminaries*

First, we introduce the definition of the local Lipschitz continuity and the locally Lipschitz function.

*Definition 2.1 (Local Lipschitz continuity [40], [43]):* A function $\hbar(\boldsymbol{x})$ is Lipschitz *near a point* $\boldsymbol{x}_0 \in \mathrm{int}\,\mathrm{dom}\hbar$ *if there exists* $K \geq 0$ *such that* $\hbar(\boldsymbol{x}) - \hbar(\boldsymbol{x}') \leq K\|\boldsymbol{x} - \boldsymbol{x}'\|, \forall \boldsymbol{x}, \boldsymbol{x}' \in B_\delta(\boldsymbol{x}_0)$ *where* $\delta > 0$ *is sufficiently small so as to have* $B_\delta(\boldsymbol{x}_0) \subset \mathrm{dom}\hbar$. *A* locally Lipschitz *function is a function that is* Lipschitz *near every point in* $\mathrm{int}\,\mathrm{dom}\hbar$.

Two important special cases of locally Lipschtz functions are continuously differentiable functions and convex functions [40]. Combining this with the boundedness of continuous functions over a compact set, it can be shown that each $\phi_j(s_j(\boldsymbol{y}_j))$ is locally Lipschitz. As a result, the objective function of problem $(P)$ is locally Lipschtiz as well. This fact will be used in establishing the optimality condition.

Next, we introduce the concept of generalized gradient which is defined for nonconvex nondifferentiable functions.

*Definition 2.2 (Generalized gradient [40], [43]):* Clarke's generalized directional derivative *of* $\hbar(\boldsymbol{x})$ *at* $\boldsymbol{x}_0$ *in the direction* $\boldsymbol{d}$, *denoted as* $\hbar^o(\boldsymbol{x}_0; \boldsymbol{d})$, *is defined by*

$$\hbar^o(\boldsymbol{x}_0; \boldsymbol{d}) = \limsup_{\substack{\boldsymbol{u} \to \boldsymbol{0} \\ \lambda \downarrow 0}} \frac{\hbar(\boldsymbol{x}_0 + \boldsymbol{u} + \lambda \boldsymbol{d}) - \hbar(\boldsymbol{x}_0 + \boldsymbol{u})}{\lambda}$$

$$= \lim_{\delta \downarrow 0} \sup_{\boldsymbol{u} \in \mathcal{B}_\delta(\boldsymbol{0}), \lambda \in (0, \delta)} \frac{\hbar(\boldsymbol{x}_0 + \boldsymbol{u} + \lambda \boldsymbol{d}) - \hbar(\boldsymbol{x}_0 + \boldsymbol{u})}{\lambda}$$
(2)

*Also,* Clarke's generalized subdifferential *of* $\hbar$ *at* $\boldsymbol{x}_0$ *is defined by*

$$\bar{\partial}\hbar(\boldsymbol{x}_0) = \{\boldsymbol{\xi} : \hbar^o(\boldsymbol{x}_0; \boldsymbol{d}) \geq \boldsymbol{\xi}^T \boldsymbol{d}\}.$$

*For any* $\boldsymbol{\xi} \in \bar{\partial}\hbar(\boldsymbol{x}_0)$, *we refer to it as* generalized gradient *of* $\hbar$ *at* $\boldsymbol{x}_0$.

As compared to the conventional directional derivative [44], the generalized directional derivative in (2) is defined with a new "base point", i.e., $\boldsymbol{x}_0 + \boldsymbol{u}$, for taking difference. Moreover, due to the *supreme* taken before the limit, it is shown in [40, Lemma 2.6] that the generalized directional derivative $\hbar^o(\boldsymbol{x}_0; \boldsymbol{d})$ is convex with respect to $\boldsymbol{d}$ even when $\hbar$ itself is nonconvex. Hence, by convex analysis, we have Theorem 2.1.

*Theorem 2.1:* Let $\hbar(\boldsymbol{x})$ be Lipschitz near $\boldsymbol{x}_0$ with local Lipschitz constant $K$. Then the following holds:

1) $\hbar^o(\boldsymbol{x}_0; \boldsymbol{0}) = 0$;
2) $\bar{\partial}\hbar(\boldsymbol{x}_0)$ is not empty and is a compact set;
3) $\|\boldsymbol{\xi}\| \leq K, \forall \ \boldsymbol{\xi} \in \bar{\partial}\hbar(\boldsymbol{x}_0)$;
4) $\hbar^o(\boldsymbol{x}_0; \boldsymbol{d}) = \max_{\boldsymbol{\xi} \in \bar{\partial}\hbar(\boldsymbol{x}_0)} \boldsymbol{\xi}^T \boldsymbol{d}, \ \forall \ \boldsymbol{d}$.

*Proof:* Since $\hbar(\boldsymbol{x})$ is Lipschitz near $\boldsymbol{x}_0$, we have $\hbar^o(\boldsymbol{x}_0; \boldsymbol{0}) = 0$ and $\|\boldsymbol{\xi}\| \leq K$ for all $\boldsymbol{\xi} \in \bar{\partial}\hbar(\boldsymbol{x}_0)$ by [40, Lemma 2.6]. Moreover, it is known from [40, Lemma 2.6] that $\hbar^o(\boldsymbol{x}_0; \boldsymbol{d})$ is a convex function with respect to $\boldsymbol{d}$. It follows that[1] $\bar{\partial}\hbar(\boldsymbol{x}_0) = \partial_{\boldsymbol{d}}\hbar^o(\boldsymbol{x}_0; \boldsymbol{0})$ is not empty and compact [44, Lemma 2.16 & Theorem 2.15], and moreover $\hbar^o(\boldsymbol{x}_0; \boldsymbol{d}) = \sup_{\boldsymbol{\xi} \in \bar{\partial}\hbar(\boldsymbol{x}_0)} \boldsymbol{\xi}^T \boldsymbol{d}, \ \forall \boldsymbol{d}$ [40, Theorem 2.5], implying part 4). This completes the proof. ∎

Furthermore, the following theorem establishes the connections between the generalized gradient and two classical concepts: the ordinary gradient and the subdifferential of convex analysis. The proof can be found in [40, Prop. 2.7 & 2.8].

*Theorem 2.2:* The following holds

1) If $\hbar(\boldsymbol{x})$ is continuously differentiable at $\boldsymbol{x}_0$, then $\bar{\partial}\hbar(\boldsymbol{x}_0) = \{\nabla \hbar(\boldsymbol{x}_0)\}$.
2) If $\hbar(\boldsymbol{x})$ is a convex function, then the Clarke's generalized gradient coincides with the subdifferential of $\hbar$, i.e., $\bar{\partial}\hbar(\boldsymbol{x}) = \partial \hbar(\boldsymbol{x})$.

Theorem 2.2 implies $\bar{\partial}\tilde{\phi}(\boldsymbol{y}_j) = \nabla \phi(s_j(\boldsymbol{y}_j)) \partial s_j(\boldsymbol{y}_j), \forall j$. Moreover, considering that both convex functions and continuously differentiable functions are locally Lipschtz, according to the result of the above two theorems, we can deduce that $\nabla \hbar(\boldsymbol{x}_0)$ is bounded if $\hbar(\boldsymbol{x})$ is continuously differentiable at $\boldsymbol{x}_0$, and that any subgradient of $\hbar(\boldsymbol{x})$ is also bounded if $\hbar(\boldsymbol{x})$ is convex.

---

[1]Considering that $\hbar^o(\boldsymbol{x}_0; \boldsymbol{d})$ is a convex function with respect to $\boldsymbol{d}$, we use $\partial_{\boldsymbol{d}} \hbar^o(\boldsymbol{x}_0; \boldsymbol{0})$ to denote its subdifferential evaluated at $\boldsymbol{d} = \boldsymbol{0}$.

## B. KKT Characterization Under Robinson's Condition

To describe optimality condition for nonlinear optimization, it is often required to assume that the problem satisfies some regularity conditions [22], [41]. In this paper, we use *Robinson's condition*, whose precise definition is given below. Note that we have provided in Appendix A some basics for better understanding Robinson's condition.

*Definition 2.3 (Robinson's condition [22], [41]):* Robinson's condition is satisfied at $\hat{z} \triangleq (\hat{x}, \hat{y})$ for problem $(P)$, if the following holds [41, Chap. 3]

$$\left\{ \begin{pmatrix} \nabla h(\hat{x}, \hat{y}) d_z \\ \nabla g_1(\hat{x}_i) d_{x_1} - v_1 \\ \vdots \\ \nabla g_n(\hat{x}_n) d_{x_n} - v_n \end{pmatrix} \middle| \begin{array}{c} d_x \in T_{\mathcal{X}}(\hat{x}), d_y \in \mathbb{R}^M, \\ v \in \mathbb{R}^q, v_{i,\ell} \leq 0, \\ \forall \ell \in I_i(\hat{x}_i), \forall i \end{array} \right\} = \mathbb{R}^p \times \mathbb{R}^q \quad (3)$$

where $d_z \triangleq (d_x, d_y)$, $v \triangleq (v_i)_i$, $v_{i,\ell}$ denotes the $\ell$-th element of $v_i$, $I_i(\hat{x}_i)$ is the $i$-th index set of active inequality constraints at $\hat{x}$, i.e.,

$$I_i(\hat{x}_i) \triangleq \{\ell \mid g_{i,\ell}(\hat{x}_i) = 0, \ 0 \leq \ell \leq q_i\},$$

where $g_{i,\ell}(\hat{x}_i)$ denotes the $\ell$-th component function of $g_i(\hat{x}_i)$. According to Theorem A.2 in Appendix A, when the system of constraints of problem $(P)$ satisfies Robinson's condition at point $\hat{z} \triangleq (\hat{x}, \hat{y})$, the tangent cone to the feasible set $\mathcal{Z}$ of problem $(P)$ exists and takes the following form [41, Chap. 3]

$$T_{\mathcal{Z}}(\hat{x}, \hat{y}) = \Big\{ d_z \triangleq (d_x, d_y) \mid d_x \in T_{\mathcal{X}}(\hat{x}), d_y \in \mathbb{R}^M,$$
$$\nabla h(\hat{x}, \hat{y}) d_z = 0, \nabla g_{i,\ell}(\hat{x}_i)^T d_{x_i} \leq 0, \ell \in I_i(\hat{x}_i), \forall i \Big\} \quad (4)$$

where $d_{x_i} \in \mathbb{R}^{n_i}$ is the $i$-th subvector of $d_x$ with $d_x = (d_{x_i})_i$.

Now we are ready to establish the KKT condition for problem $(P)$. Our proof is extended from Theorem 3.25 in [41] which deals with the case where the objective function is differentiable. Here we deal with the possibly nonconvex and nondifferentiable objective function of problem $(P)$ by using the notion of generalized directional derivative/gradient.

*Theorem 2.3:* Let $(\hat{x}, \hat{y})$ be a local minimum of problem $(P)$. Assume that Robinson's condition holds for problem $(P)$ at $(\hat{x}, \hat{y})$. Then there exist multipliers $\hat{\mu} \in \mathbb{R}^p$ and $\hat{\nu}_i \in \mathbb{R}^{q_i}$, $i = 1, 2, \ldots, n$, such that the following generalized KKT system is satisfied

$$\big(\nabla_{x_i} f(\hat{x}, \hat{y}) + \nabla_{x_i} h(\hat{x}, \hat{y})^T \hat{\mu} + \nabla_{x_i} g_i(\hat{x}_i)^T \hat{\nu}_i\big)^T$$
$$\times (x_i - \hat{x}_i) \geq 0, \ \forall \ x_i \in \mathcal{X}_i, \quad (5a)$$
$$0 \in \bar{\partial} \tilde{\phi}(y_j) + \nabla_{y_j} f(\hat{x}, \hat{y}) + \nabla_{y_j} h(\hat{x}, \hat{y})^T \hat{\mu}, \ \forall \ j, \quad (5b)$$
$$(\hat{\nu}_i)^T g_i(\hat{x}_i) = 0, \ \forall i, \quad (5c)$$
$$g_i(\hat{x}_i) \leq 0, \ \forall \ i, \quad (5d)$$
$$\hat{\nu}_i \geq 0, \ \forall \ i, \quad (5e)$$
$$h(\hat{x}, \hat{y}) = 0. \quad (5f)$$

*Proof:* The proof is divided into two steps. We first establish a necessary optimality condition, which is then shown to be equivalent to the KKT system.

**Step 1.** Recall that $F(x, y) \triangleq f(x, y) + \sum_{j=1}^{n_y} \phi_j(s_j(y_j))$ is the objective function of problem $(P)$ (5). In the first step, we show that a local optimal solution point $\hat{x}, \hat{y}$ must satisfy the following condition

$$F^o(\hat{x}, \hat{y}; d) \geq 0, \ \forall \ d \in T_{\mathcal{Z}}(\hat{x}, \hat{y}).$$

Assume on the contrary that there exists a direction $d \in T_{\mathcal{Z}}(\hat{x}, \hat{y})$ such that $F^o(\hat{x}, \hat{y}; d) < 0$. Because $d$ is a tangent direction, there exists a sequence $z^k \triangleq (x^k, y^k) \in \mathcal{Z}$ converging to $\hat{z} \triangleq (\hat{x}, \hat{y})$ and a sequence of nonnegative scalars $\tau^k \to 0$ as $k \to \infty$, such that [41, Def. 3.11]

$$\lim_{k \to \infty} \frac{z^k - \hat{z}}{\tau^k} = d.$$

It follows that

$$\lim_{k \to \infty} \frac{w^k}{\|w^k\|} = \frac{d}{\|d\|}, \text{ where } w^k \triangleq z^k - \hat{z}. \quad (6)$$

Define a sequence $\{\delta_k\}$ such that the following conditions are satisfied

$$\delta_k > \|w^k\|, \ \forall \ k \quad \text{and} \quad \lim_{k \to \infty} \delta_k \to 0. \quad (7)$$

Then we have

$$\limsup_{k \to \infty} \frac{F(z^k) - F(\hat{z})}{\|z^k - \hat{z}\|}$$
$$= \limsup_{k \to \infty} \frac{F(\hat{z} + w^k) - F(\hat{z})}{\|w^k\|}$$
$$\stackrel{(i)}{\leq} \limsup_{k \to \infty} \sup_{\substack{u \in B_{\delta_k}(0), \\ \lambda \in (0, \delta_k)}} \frac{F(\hat{z} + u + \lambda \frac{w^k}{\|w^k\|}) - F(\hat{z} + u)}{\lambda}$$
$$\stackrel{(ii)}{=} \lim_{k \to \infty} \sup_{\substack{u \in B_{\delta_k}(0), \\ \lambda \in (0, \delta_k)}} \frac{F(\hat{z} + u + \lambda \frac{w^k}{\|w^k\|}) - F(\hat{z} + u)}{\lambda}$$
$$= F^o\left(\hat{x}, \hat{y}; \frac{d}{\|d\|}\right) \stackrel{(iii)}{<} 0,$$

where (i) is due to the fact that $\|w_k\| < \delta_k$ and $0 \in B_{\delta_k}(0)$; (ii) follows from the existence of the limit, and (iii) is due to the assumption $F^o(\hat{x}, \hat{y}; d) < 0$ as well as the positive homogeneity of generalized gradient [40, Lemma 2.6], i.e., $F^o(\hat{x}, \hat{y}; \alpha d) = \alpha F^o(\hat{x}, \hat{y}; d)$ for all $\alpha \geq 0$. The above result contradicts to the fact that $(\hat{x}, \hat{y})$ is a local optimum. Hence, for any local optimum $(\hat{x}, \hat{y})$, we must have $F^o(\hat{x}, \hat{y}; d) \geq 0, \ \forall \ d \in T_{\mathcal{Z}}(\hat{x}, \hat{y})$.

**Step 2.** Based on the necessary optimality condition established in the first step, we then show that the KKT system holds for a locally optimal solution $(\hat{x}, \hat{y})$. First, by noting that $F(x, y)$ is locally Lipschitz near $(\hat{x}, \hat{y})$ (see the arguments under Definition 2.1) and using the result of Part 4) of Theorem 2.1, we have, $\exists \xi \in \bar{\partial} F(\hat{x}, \hat{y})$ such that

$$\xi^T d = F^o(\hat{x}, \hat{y}; d) \geq 0, \ \forall \ d \in T_{\mathcal{Z}}(\hat{x}, \hat{y}). \quad (8)$$

Recall the definition of polar cone (see Appendix A). Eq. (8) can be equivalently expressed as: $-\xi \in (T_{\mathcal{Z}}(\hat{x}, \hat{y}))^o$. Define

$$\mathbf{A} \triangleq \begin{pmatrix} \nabla h(\hat{x}, \hat{y}) \\ [\text{blkdiag}\{\nabla g_i(x_i)^T\}_i \ \mathbf{0}_{q \times M}] \end{pmatrix},$$
$$\mathcal{K}_1 \triangleq T_{\mathcal{X}}(\hat{x}) \times \mathbb{R}^M, \ \mathcal{K}_2 \triangleq \{0\}^p \times \mathbb{R}_-^q$$

where the notation $\text{blkdiag}\{\nabla \boldsymbol{g}_i(\boldsymbol{x}_i)^T\}_i$ denotes a $q$ by $N$ matrix which is block diagonal concatenation of matrices $\nabla \boldsymbol{g}_i(\boldsymbol{x}_i)^T$, $i = 1, 2, \ldots, n$, that is,

$$\text{blkdiag}\{\nabla \boldsymbol{g}_i(\boldsymbol{x}_i)^T\}_i$$
$$\triangleq \begin{pmatrix} \nabla \boldsymbol{g}_1(\boldsymbol{x}_1)^T & & & \\ & \nabla \boldsymbol{g}_2(\boldsymbol{x}_2)^T & & \\ & & \ddots & \\ & & & \nabla \boldsymbol{g}_n(\boldsymbol{x}_n)^T \end{pmatrix} \quad (9)$$

Assume for simplicity $I_i(\hat{\boldsymbol{x}}_i) = \{1, 2, \ldots, q_i\}$, then $T_{\mathcal{Z}}(\hat{\boldsymbol{x}}, \hat{\boldsymbol{y}})$ defined by (4) can be compactly expressed as

$$T_{\mathcal{Z}}(\hat{\boldsymbol{x}}, \hat{\boldsymbol{y}}) = \{\boldsymbol{d} \in \mathcal{K}_1 \mid \mathbf{A}\boldsymbol{d} \in \mathcal{K}_2\}.$$

Moreover, Robinson's condition (3) is equivalent to [41, pp. 102]

$$\mathbf{0} \in int\left(\{\mathbf{A}\boldsymbol{\theta} - \boldsymbol{\eta} : \boldsymbol{\theta} \in \mathcal{K}_1, \boldsymbol{\eta} \in \mathcal{K}_2\}\right).$$

It follows that (see [41, Theorem 2.36], or Theorem A.1)

$$-\boldsymbol{\xi} \in (T_{\mathcal{Z}}(\hat{\boldsymbol{x}}, \hat{\boldsymbol{y}}))^o = \mathcal{K}_1^o + \{\mathbf{A}^T \hat{\boldsymbol{\lambda}} : \hat{\boldsymbol{\lambda}} \in \mathcal{K}_2^o\} \quad (10)$$

where $\mathcal{K}_1^o = N_{\mathcal{X}}(\hat{\boldsymbol{x}}) \times \{0\}^M$ and $\mathcal{K}_2^o = \mathbb{R}^p \times \mathbb{R}_+^q$ are obtained by the definition of polar cone; see Appendix A. Eq. (10) is further equivalent to, $\exists \hat{\boldsymbol{\mu}} \in \mathbb{R}^p$, $\hat{\boldsymbol{\nu}}_i \in \mathbb{R}_+^{q_i}$, and $\hat{\boldsymbol{\lambda}} \triangleq (\hat{\boldsymbol{\mu}}, (\hat{\boldsymbol{\nu}}_i)_i)$ such that

$$-\mathbf{A}^T \hat{\boldsymbol{\lambda}} \in \mathcal{K}_1^o + \bar{\partial} F(\hat{\boldsymbol{x}}, \hat{\boldsymbol{y}}). \quad (11)$$

By using the following facts

$$\bar{\partial} F(\hat{\boldsymbol{x}}, \hat{\boldsymbol{y}}) = \prod_{i=1}^n \{(\nabla_{\boldsymbol{x}_i} f(\hat{\boldsymbol{x}}, \hat{\boldsymbol{y}}))\} \times \prod_{j=1}^{n_y} \bar{\partial} \tilde{\phi}(\boldsymbol{y}_j) \quad (12)$$

$$\mathbf{A}^T \hat{\boldsymbol{\lambda}} =$$
$$\left((\nabla_{\boldsymbol{x}_i} \boldsymbol{h}(\hat{\boldsymbol{x}}, \hat{\boldsymbol{y}})^T \hat{\boldsymbol{\mu}} + \nabla_{\boldsymbol{x}_i} \boldsymbol{g}_i(\hat{\boldsymbol{x}})^T \hat{\boldsymbol{\nu}}_i)_i, (\nabla_{\boldsymbol{y}_j} \boldsymbol{h}(\hat{\boldsymbol{x}}, \hat{\boldsymbol{y}})^T \hat{\boldsymbol{\mu}})_j\right) \quad (13)$$

$$\mathcal{K}_1^o = \prod_i^n N_{\mathcal{X}_i}(\hat{\boldsymbol{x}}) \times \prod_j^{n_y} \{0\}^{m_j} \quad (14)$$

we can recast Eq. (11) into Eqs. (5a) and (5b). Now let us take $I_i(\hat{\boldsymbol{x}}_i)$ into consideration. By the definition of polar cone, we have $\nu_{i,\ell} = 0$ for $i \notin I_i(\hat{\boldsymbol{x}}_i)$. Thus Eq. (5c) follows. The rest of equations in KKT system (5) are trivial. This completes the proof. ∎

## III. PDD METHOD AND ITS CONVERGENCE

Besides the nonconvexity and nondifferentiability, the variable coupling introduced by the equality constraint $\boldsymbol{h}(\boldsymbol{x}, \boldsymbol{y}) = \boldsymbol{0}$ further complicates problem $(P)$. Without such a coupling constraint, efficient block decomposition algorithms such as BCD, BSUM or FLEXA [45] can be applied to decompose problem $(P)$ into a sequence of small-scale problems. Unfortunately, these block decomposition methods can fail to reach any interesting solution in the presence of coupling constraint [20]. In this section we propose the PDD algorithm that relaxes the difficult coupling constraints (by using Lagrangian relaxation), performs block decomposition over the resulting augmented Lagrangian function, and utilizes appropriate penalty parameters to eventually enforce the relaxed equality constraint.

TABLE I
ALGORITHM 1: PDD METHOD FOR PROBLEM (P)

```
0.  initialize z^0 = (x^0, y^0), ϱ_0 > 0, λ_0, and set k = 1
    pick two sequences {η_k > 0}, {ε_k > 0}
1.  repeat
2.      z^k = optimize(P_{ϱ_k, λ_k}, z^{k-1}, ε_k)
3.      if ‖h(z^k)‖_∞ ≤ η_k    // case 1—AL method
4.          λ_{k+1} = λ_k + (1/ϱ_k) h(z^k)
5.          ϱ_{k+1} = ϱ_k
6.      else    // case 2—penalty method
7.          λ_{k+1} = λ_k
8.          update ϱ_{k+1} by decreasing ϱ_k
9.      end
10.     k = k + 1
11. until some termination criterion is met
```

### A. The basic PDD Method

To introduce the algorithm, denote by $\mathcal{L}(\boldsymbol{x}, \boldsymbol{y}; \boldsymbol{\lambda})$ the augmented Lagrange function with penalty parameter $\varrho$ and dual variable $\boldsymbol{\lambda}$ corresponding to the coupling constraint $\boldsymbol{h}(\boldsymbol{x}, \boldsymbol{y}) = \boldsymbol{0}$. Further, let us define an *AL problem* $(P_{\varrho, \boldsymbol{\lambda}})$ as follows

$$(P_{\varrho, \boldsymbol{\lambda}}) \quad \min_{\boldsymbol{x}_i \in \widetilde{\mathcal{X}}_i, \boldsymbol{y}} \left\{ \mathcal{L}(\boldsymbol{x}, \boldsymbol{y}; \boldsymbol{\lambda}) \triangleq f(\boldsymbol{x}, \boldsymbol{y}) + \sum_{j=1}^{n_y} \phi_j(s(\boldsymbol{y}_j)) \right.$$
$$\left. + \boldsymbol{\lambda}^T \boldsymbol{h}(\boldsymbol{x}, \boldsymbol{y}) + \frac{1}{2\varrho} \|\boldsymbol{h}(\boldsymbol{x}, \boldsymbol{y})\|^2 \right\} \quad (15)$$

where $\widetilde{\mathcal{X}}_i \triangleq \{\boldsymbol{x}_i \mid \boldsymbol{g}_i(\boldsymbol{x}_i) \leq 0, \, \boldsymbol{x}_i \in \mathcal{X}_i\}$.

The basic PDD method is a double-loop iterative algorithm, where, the inner loop approximately solves the AL subproblem (15) while the outer loop updates the dual variable or the penalty parameter if necessary. We present the PDD method in TABLE I, where the notation '$optimize(P_{\varrho_k, \boldsymbol{\lambda}_k}, \boldsymbol{z}^{k-1}, \epsilon_k)$' represents some optimization oracle employed to iteratively solve problem $(P_{\varrho_k, \boldsymbol{\lambda}_k})$ to some accuracy $\epsilon_k$ starting from the initial point $\boldsymbol{z}^{k-1}$. In particular, to fully exploit the problem structure, the typical optimization oracle could be BCD-type algorithms. That is, we can instantiate the optimization oracle in Step 2 by using the classical BCD algorithm [22], or some inexact variants of BCD, such as the BSUM [46] algorithm. Furthermore, we update the dual variable $\boldsymbol{\lambda}_k$ when the constraint violation $\|\boldsymbol{h}(\boldsymbol{z}^k)\|_\infty$ is relatively small (i.e., Step 4); otherwise we decrease the penalty parameter $\varrho_k$ (i.e., Step 8). Therefore, the PDD method adaptively switches between the AL and the penalty method. This adaptive strategy is expected to find an appropriately penalty parameter $\varrho$, with which the AL method could eventually converge. In the PDD method, the parameter $\eta_k > 0$ measures the constraint violation and the parameter $\epsilon_k > 0$ controls the accuracy of the optimization oracle, with both parameters going to zero as the number of outer iterations $k$ increases.

### B. Convergence Analysis for PDD

In the following, we address the convergence issue of the PDD method. To do so, we define $e^k$ and $\Delta_j^k$ in (16) and

$$e^k = \mathcal{P}_{\mathcal{X}}\{x^k - \nabla_x \mathcal{L}_k(x^k, y^k) - \nabla g(x^k)^T \nu^k\} - x^k, \tag{16}$$

$$\Delta_j^k = y_j^k - \arg\min_{y_j} \left\{ \begin{array}{c} \phi_j'(s_j(y_j^k))s_j(y_j) + \frac{1}{2}\|y_j - y_j^k\|^2 \\ + \left(\nabla_{y_j} f(x^k, y^k) + \nabla_{y_j} h(x^k, y^k)^T \left(\frac{1}{\varrho_k}h(x^k, y^k) + \lambda^k\right)\right)^T (y_j - y_j^k) \end{array} \right\} \tag{17}$$

(17) (see the top of the next page), where $g(x) \triangleq (g_i(x_i))_i$. We will show that, when these two terms go to zero, the first order optimality condition of the AL problem with respect to $x$ and $y$ holds true. The main convergence result is presented in Theorem 3.1.

*Theorem 3.1:* Let $\{x^k, y^k, \nu^k\}$ be the sequence generated by Algorithm 1 for problem $(P)$, where $\nu^k = (\nu_i^k)_i$ denotes the Lagrange multipliers associated with the constraints $g_i(x_i) \leq 0, \forall i$. The termination condition for the optimization oracle involved in Algorithm 1 is

$$\max\left(\|e^k\|_\infty, \|\Delta^k\|_\infty\right) \leq \epsilon_k, \forall k \tag{18}$$

with $\epsilon_k, \eta_k, \varrho_k \to 0$ as $k \to \infty$. Suppose that $(x^*, y^*)$ is a limit point of the sequence $\{x^k, y^k\}$ and at the limit point $(x^*, y^*)$ the Robinson's condition holds for problem $(P)$. Then $(x^*, y^*)$ satisfies $h(x^*, y^*) = 0$, and it is a KKT point of problem $(P)$ that satisfies (5).

*Proof:* Our proof consists of two steps, in the first step we will utilize Robinson's condition to argue that $\{\mu^k\}$ (cf. (21)) is a bounded sequence. Then based on this result we will argue that the sequence converges to KKT points.

**Step 1.** First, we show that a key inequality [see (25)] holds for $\{(x^k, y^k)\}$. Without loss of generality, we assume that the sequence $\{(x^k, y^k)\}$ converges to $(x^*, y^*)$ (otherwise we can restrict to a convergent subsequence of $\{(x^k, y^k)\}$). By noting that $\mathcal{X}$ is a closed convex set, we have $x^* \in \mathcal{X}$. By the definition of $e^k$ and using *projection theorem* [22, Prop. 2.1.3 (b)], we have

$$\left(x - (x^k + e^k)\right)^T \left((x^k - \nabla_x \mathcal{L}_k(x^k, y^k) - \nabla g(x^k)^T \nu^k) - (x^k + e^k)\right) \leq 0, \forall x \in \mathcal{X}, \forall k. \tag{19}$$

It follows that

$$-\left(x - (x^k + e^k)\right)^T \left(\nabla_x \mathcal{L}_k(x^k, y^k) + \nabla g(x^k)^T \nu^k + e^k\right) \leq 0, \forall x \in \mathcal{X}, \forall k. \tag{20}$$

Let us define a "virtual" multiplier vector as

$$\mu^k \triangleq \frac{1}{\varrho_k} h(x^k, y^k) + \lambda_k. \tag{21}$$

Then we have

$$\nabla_x \mathcal{L}_k(x^k, y^k) = \nabla_x f(x^k, y^k) + \nabla_x h(x^k, y^k)^T \mu^k.$$

Plugging the above equality into (20), we obtain

$$-\left(x - (x^k + e^k)\right)^T \left(\nabla_x f(x^k, y^k) + \nabla_x h(x^k, y^k)^T \mu^k + \nabla g(x^k)^T \nu^k + e^k\right) \leq 0, \forall x \in \mathcal{X}, \forall k. \tag{22}$$

On the other hand, by the definition of $\Delta_j^k$ and (21), we have that for all $j$ the following identity holds

$$y_j^k - \Delta_j^k = \arg\min_{y_j} \left\{ \phi_j'(s_j(y_j^k))s_j(y_j) + \frac{1}{2}\|y_j - y_j^k\|^2 \right.$$
$$\left. + \left(\nabla_{y_j} f(x^k, y^k) + \nabla_{y_j} h(x^k, y^k)^T \mu^k\right)^T (y_j - y_j^k) \right\} \tag{23}$$

By the optimality condition of the above problem, we have, $\exists \xi_j^k \in \phi_j'(s_j(y_j^k))\partial s_j(y_j^k - \Delta_j^k), \forall j$ such that

$$\sum_{j=1}^{n_y} \left(\xi_j^k - \Delta_j^k + \nabla_{y_j} f(x^k, y^k) + \nabla_{y_j} h(x^k, y^k)^T \mu^k\right)^T$$
$$\times (y_j - y_j^k + \Delta_j^k)) = 0. \tag{24}$$

Combining (24) with (22), we have

$$(\nabla f(x^k, y^k) + \chi^k + \nabla h(x^k, y^k)^T \mu^k)^T$$
$$\times (x - x^k - e^k, y - y^k + \Delta^k) \geq 0, \forall x \in \mathcal{X}, y \in \mathbb{R}^M. \tag{25}$$

where

$$\chi^k \triangleq \left\{ \begin{array}{c} \nabla g(x^k)^T \nu^k + e^k \\ \xi^k + \Delta^k \end{array} \right\} \tag{26}$$
$$\Delta^k \triangleq (\Delta_j^k)_j, \xi = (\xi_j^k)_j.$$

Next, we prove that $\mu^k$ is bounded by contradiction and using Robinson condition. Assume, to the contrary, that $\mu^k$ is unbounded. Define $\bar{\mu}^k \triangleq \frac{\mu^k}{\|\mu^k\|}$. Since $\{\bar{\mu}^k\}$ is bounded, there must exist a convergent subsequence $\{\bar{\mu}^{k_r}\}$. Let $\mu^{k_r} \to \bar{\mu}$ as $r \to \infty$. On the other hand, since $f(x, y)$ and $g(x)$ are continuously differentiable, $\nabla f(x^k, y^k)$ and $\nabla g(x^k)$ are bounded. Moreover, by Theorem 2.1, we know that $\xi^k$ is bounded. Also, by Robinson's condition and Lemma 3.26 in [41], we conclude that $\nu^k$ is bounded. As a result, $\chi^k$ is bounded[2]. By dividing both sides of (25) by $\|\mu^k\|$ and using the boundedness of $\nabla f(x^k, y^k)$ and $\chi^k$, we have for sufficiently large $r$

$$-\left(x - (x^{k_r} + e^{k_r}), y - (y^{k_r} - \Delta^{k_r})\right)^T$$
$$\times \left(\nabla h(x^{k_r}, y^{k_r})^T \bar{\mu}^{k_r}\right) \leq 0, \forall x \in \mathcal{X}. \tag{27}$$

Note that $\nabla h(x, y)$ is continuous in $(x, y)$. Moreover, by assumption

$$\max\left(\|e^k\|_\infty, \|\Delta^k\|_\infty\right) \leq \epsilon_k, \forall k, \tag{28}$$

we have $e^k \to 0$ and $\Delta^k \to 0$ due to $\epsilon_k \to 0$ as $k \to 0$. In addition, it holds that $(x^{k_r}, y^{k_r}) \to (x^*, y^*)$ and $\mu^{k_r} \to \bar{\mu}$ as $r \to \infty$. Hence, taking limits on both sides of (25), we have

$$-(x - x^*, y - y^*)^T \nabla h(x^*, y^*)^T \bar{\mu} \leq 0, \forall x \in \mathcal{X}, y \in \mathbb{R}^M. \tag{29}$$

---
[2]Note that the objective function of problem $(P_{\varrho, \lambda})$ is continuously differentiable in $x$. Thus we can apply here Lemma 3.26 in [41].

Utilizing the first part of the Robinson's condition, that is
$$\{\nabla h(x^*, y^*)(d_x, d_y) : d_x \in T_\mathcal{X}(x^*), d_y \in \mathbb{R}^M\} = \mathbb{R}^p, \quad (30)$$
it follows that there exists some $x \in \mathcal{X}$, $y \in \mathbb{R}^M$ and $c > 0$ such that $-\bar{\mu} = c\nabla h(x^*, y^*)(x - x^*, y - y^*)$. This together with (29) implies $\bar{\mu} = 0$, contradicting the identity $\|\bar{\mu}\| = 1$. Hence, $\{\mu^k\}$ is bounded.

**Step 2.** Next we show that the algorithm indeed reaches the KKT points. From Steps 3-9, we observe that, either both $\{\mu^k\}$ and $\{\lambda_k\}$ are bounded with $\varrho_k \to 0$ (i.e., case 2 in Algorithm 1), or $\mu^k - \lambda_k \to 0$ with $\varrho_k$ bounded (i.e., case 1 in Algorithm 1). Hence, from the definition (21) we must have
$$h(x^k, y^k) = \varrho_k(\mu^k - \lambda_k) \to 0.$$
which implies that $h(x^*, y^*) = 0$. That is, the equality constraint will be satisfied in the limit. In addition, due to the boundedness of $\{\mu^k\}$, there exists a convergent subsequence $\{\mu^{k_r}\}$ that we assume converge to $\mu^*$. By restricting to the subsequence $\{\mu^{k_r}\}$ and taking limits on both sides of (22), we have
$$(x - x^*)^T \left(\nabla_x f(x^*, y^*) + \nabla_x h(x^*, y^*)^T \mu^* \right.$$
$$\left. + \nabla g(x^*)^T \nu^* \right) \geq 0, \forall x \in \mathcal{X}, \quad (31)$$

On the other hand, since problem (23) has a unique solution, by restricting to a convergent subsequence, we can take limit on both sides of (23), leading to
$$y_j^* = \arg\min_{y_j} \phi_j'(s_j(y_j^*))s_j(y_j) + \frac{1}{2}\|y_j - y_j^*\|^2 \quad (32)$$
$$+ \left(\nabla_{y_j} f(x^*, y^*) + \nabla_{y_j} h(x^*, y^*)^T \mu^*\right)^T (y_j - y_j^*), \forall j.$$
It follows that
$$0 \in \phi_j'(s_j(y_j^*))\partial s_j(y_j^*) + \nabla_{y_j} f(x^*, y^*) + \nabla_{y_j} h(x^*, y^*)^T \mu^*, \forall j \quad (33)$$
In addition, $g(x^k) \leq 0$ implies $g(x^*) \leq 0$. Moreover, since $\nu^k$ are the Lagrange multiplier associated with the constraints $g(x) \leq 0$, we have $g(x^k)^T \nu^k = 0$ and $\nu^k \geq 0$. It follows that
$$g(x^*)^T \nu^* = 0 \text{ and } \nu^* \geq 0. \quad (34)$$
Combining Eqs. (31), (33), (34) and the fact $h(x^*, y^*) = 0$, $g(x^*) \leq 0$, and $x^* \in \mathcal{X}$, we conclude that $(x^*, y^*)$ satisfies the KKT condition of problem $(P)$. This completes the proof. ∎

*Remark 3.1:* We note that in the above proof, the Robinson's condition has been used in a slightly different way than in the proof of Theorem 2.3. In particular, in Theorem 2.3, the condition is assumed on a local minimizer $(\hat{x}, \hat{y})$, which is obviously a *feasible* solution for problem (P). On the other hand, in Theorem 3.1, the Robinson's condition is assumed on a limit point $(x^*, y^*)$ generated by the PDD algorithm, and such a point may not be feasible for the constraints $h(x^*, y^*) = 0$ to begin with. Therefore, in practical applications, in order to use Theorem 3.1, one has to check whether Robinson's condition holds for all $(x, y)$ satisfying the constraints that $x \in \mathcal{X}$, $g_i(y) \leq 0$, $\forall i$ (but not necessarily satisfying $h(x, y) = 0$). This will be done for each application that we will study in Part II of this paper.

TABLE II
ALGORITHM 2: PDD ALGORITHM FOR PROBLEM (P)

---
0. initialize $z^0 = (x^0, y^0)$, $\varrho_0 > 0$, $\lambda_0$, and $k = 1$
   pick two sequences $\{\eta_k > 0\}$, $\{\epsilon_k > 0\}$
1. **repeat**
2.     $z^k = \text{optimize}(P_{\varrho_k, \lambda_k}, z^{k-1}, \epsilon_k)$
4.     $\lambda_{k+1} = \lambda_k + \frac{1}{\varrho_k} h(z^k)$
5.     update $\varrho_{k+1}$ by decreasing $\varrho_k$
6.     $k = k + 1$
7. **until** some termination criterion is met
---

## C. PDD Method With Increasing Penalty

The basic PDD method is expected to be able to achieve convergence with *finite* penalty for many practical applications. However, it requires frequent evaluation of constraint violation, an operation that can be costly for certain applications. To overcome this weakness, we propose a simple variant of the basic PDD method; see TABLE II for the detailed description. The main difference lies in that we always keep *increasing* the penalty and updating the dual variable. Hence this variant is referred to as *increasing penalty dual decomposition* (IPDD) method. The following theorem shows that every limit point of the iterates generated by the IPDD is a KKT point of problem $(P)$ under Robinson's condition.

*Theorem 3.2:* Let $\{x^k, y^k, \nu^k\}$ be the sequence generated by Algorithm 2 for problem $(P)$, where $\nu^k = (\nu_i^k)_i$ denote the Lagrange multipliers associated with the constraints $g_i(x_i) \leq 0, \forall i$. The termination condition for the optimization oracle involved in Algorithm 2 is given in (18) with $\epsilon_k, \eta_k, \varrho_k \to 0$ as $k \to \infty$. Suppose that $(x^*, y^*)$ is a limit point of the sequence $\{x^k, y^k\}$ and the condition (30) holds at $(x^*, y^*)$, then the point $(x^*, y^*)$ satisfies $h(x^*, y^*) = 0$. Furthermore, suppose that Robinson's condition holds for problem $(P)$ at $(x^*, y^*)$. Then $(x^*, y^*)$ is a KKT point of problem $(P)$, i.e., it satisfies the KKT system (5) of problem $(P)$.

*Proof:* Following the same argument as that of the proof of Theorem 3.1, we can prove 1) all the KKT equations except $h(x^*, y^*) = 0$ and 2) that the sequence $\{\mu^k\}$ is bounded. By checking the definition of $\mu^k$ and the dual update in Step 4 of Algorithm 2, we have $\lambda_{k+1} = \mu^k$. It follows that the sequence $\{\lambda^k\}$ is bounded, implying $\|\lambda_{k+1} - \lambda_k\|$ is bounded. Since it holds that $\varrho_k \to 0$ as $k \to \infty$, we have from the dual update that $\|h(x^k, y^k)\| = \varrho_k\|\lambda_{k+1} - \lambda_k\| \to 0$ as $k \to \infty$, implying $h(x^*, y^*) = 0$. This completes the proof. ∎

## IV. RANDOMIZED BSUM FOR PROBLEM $(P_{\varrho_k, \lambda_k})$

In the PDD/IPDD method, BCD-type algorithms are typically used as optimization oracles in Step 2 to solve problem $(P_{\varrho_k, \lambda_k})$, and it is assumed to be able to guarantee Eq. (18). However, the convergence theory of the basic BSUM algorithm [46] (which includes the exact BCD method [22] as a special case) is established only for convex constraint cases. By considering a random block update rule, we here provide an extension of the basic BSUM algorithm, termed rBSUM, which is applicable for problems with nonconvex constraints. In the following, we present the rBSUM algorithm with a convergence analysis. In particular, we show that the proposed rBSUM can reach KKT solutions of problem $(P_{\varrho_k,\boldsymbol{\lambda}_k})$, therefore ensuring Eq. (18).

To proceed, we define $\boldsymbol{z} = (\boldsymbol{z}_i)_i$ with $\boldsymbol{z}_i = \boldsymbol{x}_i$ for $i = 1, 2, \ldots, n$ and $\boldsymbol{z}_{n+j} = \boldsymbol{y}_j$ for $j = 1, 2, \ldots, n_y$, i.e., $\boldsymbol{z} = (\boldsymbol{x}, \boldsymbol{y})$. Furthermore, for notational simplicity, we omit $k$ for problem $(P_{\varrho_k,\boldsymbol{\lambda}_k})$ and denote its objective function simply as $\mathcal{L}(\boldsymbol{z})$. Thus, let us consider the rBSUM algorithm for solving

$$\begin{aligned}\min_{\boldsymbol{z}} \quad & \mathcal{L}(\boldsymbol{z}_1, \boldsymbol{z}_2, \ldots, \boldsymbol{z}_{n_z}) \\ \text{s.t.} \quad & \boldsymbol{z}_i \in \mathcal{X}_i, i = 1, 2, \ldots, n, \\ & \boldsymbol{g}_i(\boldsymbol{z}_i) \leq 0, i = 1, 2, \ldots, n\end{aligned} \quad (35)$$

where $n_z = n + n_y$ is the total number of block variables. At each iteration, the rBSUM updates one block variable by minimizing a locally tight upper bound $u_i(\cdot; \cdot)$ of the objective function, while fixing the rest of the blocks. Let $\tilde{\mathcal{X}}_{i+n} = \mathbb{R}^{m_i}$, $i = 1, 2, \ldots, n_y$ and define $\tilde{\mathcal{X}} \triangleq \tilde{\mathcal{X}}_1 \times \tilde{\mathcal{X}}_2 \times \ldots \times \tilde{\mathcal{X}}_{n_z}$. The rBSUM algorithm is summarized in TABLE III, where Steps 3 and 4 generate a random index set $\mathcal{I}$ specifying the update order of block variables. In what follows, we study the convergence of the rBSUM algorithm.

First, we make the following assumption on $u_i(\cdot; \cdot)$.

*Assumption 4.1:*

$$u_i(\boldsymbol{z}_i; \boldsymbol{z}) = \mathcal{L}(\boldsymbol{z}), \forall \boldsymbol{z} \in \tilde{\mathcal{X}}, \forall i; \quad (36a)$$

$$u_i(\boldsymbol{v}_i; \boldsymbol{z}) \geq \mathcal{L}(\boldsymbol{z}_{<i}, \boldsymbol{v}_i, \boldsymbol{z}_{>i}), \forall \boldsymbol{v}_i \in \tilde{\mathcal{X}}_i, \forall \boldsymbol{z} \in \tilde{\mathcal{X}}, \forall i; \quad (36b)$$

$$u_i^o(\boldsymbol{v}_i; \boldsymbol{z}, \boldsymbol{d}_i)|_{\boldsymbol{v}_i=\boldsymbol{z}_i} = \mathcal{L}^o(\boldsymbol{z}; \boldsymbol{d}), \forall \boldsymbol{d} = (\boldsymbol{0}, \ldots, \boldsymbol{0}, \boldsymbol{d}_i, \boldsymbol{0}, \ldots, \boldsymbol{0})$$
$$\text{s.t. } \boldsymbol{x}_i + \boldsymbol{d}_i \in \tilde{\mathcal{X}}_i, \forall i; \quad (36c)$$

$$u_i(\boldsymbol{v}_i; \boldsymbol{z}) \text{ is continuous in } (\boldsymbol{v}_i, \boldsymbol{z}), \forall i. \quad (36d)$$

In the above assumption, $\boldsymbol{v}_i$ is the $i$-th block component of $\boldsymbol{v}$, having the same size as $\boldsymbol{z}_i$; the notations $\boldsymbol{z}_{<i}$ and $\boldsymbol{z}_{>i}$ represent the block components of $\boldsymbol{z}$ with their indices less than $i$ or larger than $i$, respectively; $u_i^o(\boldsymbol{v}_i; \boldsymbol{z}, \boldsymbol{d}_i)$ denotes the generalized directional derivative of $u_i(\cdot; \boldsymbol{z})$ with respect to $\boldsymbol{v}_i$ along the direction $\boldsymbol{d}_i$; and $\mathcal{L}^o(\boldsymbol{z}; \boldsymbol{d})$ denotes the generalized directional derivative of $\mathcal{L}(\cdot)$ with respect to $\boldsymbol{z}$ along the direction $\boldsymbol{d}$. The assumption (36c) guarantees that the first order behavior of $u_i(\cdot, \boldsymbol{z})$ is the same as $\mathcal{L}(\cdot)$ locally [46], hence it is referred to as the gradient consistency assumption.

Second, we give the definition of regular functions which will be used later.

*Definition 4.1 (Regularity of a function):* A function $\hbar(\cdot)$ is *regular* at $\boldsymbol{x} = (\boldsymbol{x}_i)_i$ if the following implication holds

$$\hbar^o(\boldsymbol{x}; \boldsymbol{d}) \geq 0, \forall \boldsymbol{d} = (\boldsymbol{d}_i)_i \iff \hbar^o(\boldsymbol{x}; \boldsymbol{d}_i^0) \geq 0,$$
$$\forall \boldsymbol{d}_i^0 \triangleq (\boldsymbol{0}, \ldots, \boldsymbol{0}, \boldsymbol{d}_i, \boldsymbol{0}, \ldots, \boldsymbol{0}), \forall i.$$

Based on the above assumption and the definition of regular functions, we next prove that, with probability one (w.p.1.) the sequence generated by the rBSUM algorithm converges to the set of stationary/KKT solutions of problem (35).

*Theorem 4.1:* Let Assumption 4.1 holds. Furthermore, assume that $\mathcal{L}(\cdot)$ is bounded below in $\tilde{\mathcal{X}}$ and it is regular at every point in $\tilde{\mathcal{X}}$. Then every limit point of the iterates generated by the rBSUM algorithm is a stationary point of problem (35)

TABLE III
ALGORITHM 3: sBSUM ALGORITHM

| |
|---|
| 0.   initialize $\boldsymbol{z}^0 \in \tilde{\mathcal{X}}$ and set $k = 0$ |
| 1.   **repeat** |
| 2.     $\boldsymbol{w} = \boldsymbol{z}^k$ |
| 3.     uniformly randomly pick $i_k \in \{1, \ldots, n_z\}$ |
| 4.     $\mathcal{I}_k = \{i_k, 1, 2, \ldots, i_k - 1, i_k + 1, \ldots, n_z\}$ |
| 5.     **for** each $i \in \mathcal{I}_k$ |
| 6.       $\mathcal{A}_i^k = \arg\min_{\boldsymbol{z}_i \in \tilde{\mathcal{X}}_i} u_i(\boldsymbol{z}_i; \boldsymbol{w})$ |
| 7.       set $\boldsymbol{w}_i$ to be an arbitrary element in $\mathcal{A}_i^k$ |
| 8.     **end** |
| 9.     $\boldsymbol{z}^{k+1} = \boldsymbol{w}$ |
| 10.    $k = k + 1$ |
| 11. **until** some termination criterion is met |

w.p.1. Moreover, if Robinson's condition holds for problem (35) at the limit point, then the limit point is also a KKT point of problem (35).

*Proof:* It is easily seen that Steps 3 and 4 can generate $n_z$ permutations of the index set in total. Let $\pi$ denote the index of permutation and $\pi(1)$ denote the first number of the $\pi$-th permutation. Moreover, let $q_\pi > 0$ denote the probability of permutation $\pi$, with $\sum_{\pi=1}^{n_z} q_\pi = 1$. Then, we have

$$\mathbb{E}[\mathcal{L}(\boldsymbol{z}^{k+1}) \mid \boldsymbol{z}^k] = \sum_{\pi=1}^{n_z} q_\pi \mathcal{L}(\boldsymbol{z}^{\pi,k+1}) \quad (37)$$

where $\boldsymbol{z}^{\pi,k+1}$ denotes the update obtained by running one iteration of rBSUM (given $\boldsymbol{z}^k$) according to the block selection rule specified by the $\pi$-th permutation. Due to the upper bound assumption (36b) and the update rule, it must hold that

$$\mathcal{L}(\boldsymbol{z}^{\pi,k+1}) \leq \min_{\boldsymbol{z}_{\pi(1)} \in \tilde{\mathcal{X}}_{\pi(1)}} u_{\pi(1)}(\boldsymbol{z}_{\pi(1)}; \boldsymbol{z}^k), \forall \pi. \quad (38)$$

Combining (37) and (38), we have

$$\mathbb{E}[\mathcal{L}(\boldsymbol{z}^{k+1}) \mid \boldsymbol{z}^k] \leq \mathcal{L}(\boldsymbol{z}^k) - \sum_{\pi=1}^{n_z} q_\pi \bigg(\mathcal{L}(\boldsymbol{z}^k) - \min_{\boldsymbol{z}_{\pi(1)} \in \tilde{\mathcal{X}}_{\pi(1)}} u_{\pi(1)}(\boldsymbol{z}_{\pi(1)}; \boldsymbol{z}^k)\bigg) \quad (39)$$

which implies that $\mathcal{L}(\boldsymbol{z}^k)$ is a supermartingale and thus converges [47], and moreover the following holds w.p.1.,

$$\sum_{k=1}^\infty \sum_{\pi=1}^{n_z} q_\pi \bigg(\mathcal{L}(\boldsymbol{z}^k) - \min_{\boldsymbol{z}_{\pi(1)} \in \tilde{\mathcal{X}}_{\pi(1)}} u_{\pi(1)}(\boldsymbol{z}_{\pi(1)}; \boldsymbol{z}^k)\bigg) < \infty \quad (40)$$

as $\mathcal{L}(\cdot)$ is bounded from below. Thus, by noting $\mathcal{L}(\boldsymbol{z}^k) \geq \min_{\boldsymbol{z}_{\pi(1)} \in \tilde{\mathcal{X}}_{\pi(1)}} u_{\pi(1)}(\boldsymbol{z}_{\pi(1)}; \boldsymbol{z}^k), \forall \pi$, we must have, w.p.1.,

$$\lim_{k \to \infty} \bigg(\mathcal{L}(\boldsymbol{z}^k) - \min_{\boldsymbol{z}_{\pi(1)} \in \tilde{\mathcal{X}}_{\pi(1)}} u_{\pi(1)}(\boldsymbol{z}_{\pi(1)}; \boldsymbol{z}^k)\bigg) = 0, \forall \pi. \quad (41)$$

Now let us restrict our analysis to a convergent subsequence $\{\boldsymbol{z}^{k_j}\}$ with $\lim_{j\to\infty} \boldsymbol{z}^{k_j} = \boldsymbol{z}^\infty$. We have from (41) and the continuity of $\mathcal{L}(\cdot)$ that

$$\lim_{j \to \infty} \min_{\boldsymbol{z}_{\pi(1)} \in \tilde{\mathcal{X}}_{\pi(1)}} u_{\pi(1)}(\boldsymbol{z}_{\pi(1)}; \boldsymbol{z}^{k_j}) = \mathcal{L}(\boldsymbol{z}^\infty), \forall \pi, \text{ w.p.1.} \quad (42)$$

On the other hand, according to the update rule, we have

$$\min_{\boldsymbol{z}_{\pi(1)} \in \tilde{\mathcal{X}}_{\pi(1)}} u_{\pi(1)}(\boldsymbol{z}_{\pi(1)}; \boldsymbol{z}^{k_j}) \leq u_{\pi(1)}(\boldsymbol{z}_{\pi(1)}; \boldsymbol{z}^{k_j}),$$
$$\forall \boldsymbol{z}_{\pi(1)} \in \tilde{\mathcal{X}}_{\pi(1)}, \forall \pi, \text{ w.p.1.} \quad (43)$$

By taking limit as $j \to \infty$ on both sides of (43), and using (42) and the continuity of $u_i(\cdot;\cdot)$, we obtain

$$\mathcal{L}(\boldsymbol{z}^\infty) \leq u_{\pi(1)}(\boldsymbol{z}_{\pi(1)}; \boldsymbol{z}^\infty), \forall \boldsymbol{z}_{\pi(1)} \in \tilde{\mathcal{X}}_{\pi(1)}, \forall \pi, \text{ w.p.1.} \quad (44)$$

Due to the function value consistency assumption (36a), we have $\mathcal{L}(\boldsymbol{z}^\infty) = u(\boldsymbol{z}_i^\infty; \boldsymbol{z}^\infty), \forall i$, and thus

$$u(\boldsymbol{z}_{\pi(1)}^\infty; \boldsymbol{z}^\infty) \leq u_{\pi(1)}(\boldsymbol{z}_{\pi(1)}; \boldsymbol{z}^\infty), \forall \boldsymbol{z}_{\pi(1)} \in \tilde{\mathcal{X}}_{\pi(1)}, \forall \pi, \text{ w.p.1.} \quad (45)$$

Note that the above inequality holds for all permutations. Therefore, we have that w.p.1.,

$$u_i(\boldsymbol{z}_i^\infty; \boldsymbol{z}^\infty) \leq u_i(\boldsymbol{z}_i; \boldsymbol{z}^\infty), \forall \boldsymbol{z}_i \in \tilde{\mathcal{X}}_i, \forall i. \quad (46)$$

It follows that[3]

$$u_i^o(\boldsymbol{z}_i^\infty; \boldsymbol{z}^\infty, \boldsymbol{d}_i) \geq 0, \forall \boldsymbol{d}_i \in T_{\tilde{\mathcal{X}}_i}(\boldsymbol{z}_i^\infty), \forall i. \quad (47)$$

where

$$T_{\tilde{\mathcal{X}}_i}(\boldsymbol{z}_i^\infty) = \begin{cases} \left\{ \boldsymbol{d}_i \mid \boldsymbol{d}_i \in T_{\mathcal{X}_i}(\boldsymbol{z}_i^\infty), \\ \nabla \boldsymbol{g}_{i,\ell}(\boldsymbol{z}_i^\infty)^T \boldsymbol{d}_i \leq 0, \ell \in I_i(\boldsymbol{z}_i^\infty) \right\}, 1 \leq i \leq n \\ \mathbb{R}^{m_i - n}, i = n+1, \ldots, n_z. \end{cases} \quad (48)$$

Thus, by the gradient consistency assumption (36c), we have from (47) that w.p.1.,

$$\mathcal{L}^o(\boldsymbol{z}^\infty; \boldsymbol{d}_i^0) \geq 0, \forall \boldsymbol{d}_i^0 \triangleq (\boldsymbol{0}, \ldots, \boldsymbol{0}, \boldsymbol{d}_i, \boldsymbol{0}, \ldots, \boldsymbol{0}), \quad (49)$$
$$\boldsymbol{d}_i \in T_{\tilde{\mathcal{X}}_i}(\boldsymbol{z}_i^\infty), \forall i.$$

Since $\mathcal{L}(\boldsymbol{z})$ is regular at $\boldsymbol{z}^\infty$, it follows that w.p.1.,

$$\mathcal{L}^o(\boldsymbol{z}^\infty; \boldsymbol{d}) \geq 0, \forall \boldsymbol{d} = (\boldsymbol{d}_1, \boldsymbol{d}_2, \ldots, \boldsymbol{d}_{n_z})$$
$$\text{with } \boldsymbol{d}_i \in T_{\tilde{\mathcal{X}}_i}(\boldsymbol{z}_i^\infty), i = 1, 2, \ldots, n_z. \quad (50)$$

By applying to Eq. (50) a similar argument as that for the second part of proof of Theorem 2.3, we can show under Robinson's condition that, there exists multipliers $(\hat{\boldsymbol{\nu}}_j)_j$ associated with the inequality constraints such that, the KKT condition of problem (35), i.e., Eqs. (5a-5e) with $(\hat{\boldsymbol{x}}, \hat{\boldsymbol{y}}) = \boldsymbol{z}^\infty$, holds true w.p.1.. This completes the proof. ∎

Since the rBSUM algorithm achieves convergence to KKT points *w.p.1* under Robinson's condition, we modify the claims of Theorem 3.1 & 3.2 for the case where rBSUM is used as the optimization oracle.

*Corollary 4.1:* Suppose that the parameter settings and the termination conditions of the optimization oracle in Theorem 3.1 & 3.2 are used, and that rBSUM is used as the optimization oracle. Then every limit point of the sequence generated by the PDD/IPDD method is a KKT point w.p.1 when rBSUM is used as the optimization oracle, provided that the Robinson's condition is satisfied at the limit point.

---

[3]This can be proven following a similar argument as that for the first part of proof of Theorem 2.3.

## V. Discussion

### A. The Robinson's condition

It is well-known that constraint qualification (CQ) conditions (or regularity conditions) are often needed to precisely describe the first-order optimality condition for nonlinear optimization. In our KKT and convergence analysis, Robinson's condition is assumed as a type of CQ. Similarly to many other CQs, such condition is generally difficult to check, but it is a standard one and has been used in many existing works on constrained optimization, e.g., [23], [41], [48], [49]. For ease of understanding Robinson's condition, [41, Lemma 3.16] has provided a simple sufficient condition. That is, *if the rows of $\nabla \boldsymbol{h}(\boldsymbol{z}^*)$ are linearly independent and moreover there exists $\boldsymbol{z}^{int} = (\boldsymbol{x}^{int}, \boldsymbol{y}^{int}) \in int(\mathcal{X} \times \mathbb{R}^M)$ such that $\nabla \boldsymbol{h}(\boldsymbol{z}^*)(\boldsymbol{z}^{int} - \boldsymbol{z}^*) = \boldsymbol{0}$ and $\nabla g_{i\ell}(\boldsymbol{x}_i^*)(\boldsymbol{x}_i^{int} - \boldsymbol{x}_i^*) < 0$, $\forall \ell \in I_i(\boldsymbol{x}_i^*), \forall i$, then Robinson's condition (3) holds true.*

Below, we summarize the relationship between the Robinson's condition and a few commonly used CQs.

*1) MFCQ:* When $\mathcal{X} = \mathbb{R}^N$, the above sufficient condition for Ronbinson's condition reduces to the well-known *Mangasarian-Fromovitz constraint qualification* (MFCQ). Moreover, it is shown in [41, Lemma 3.17] that Robinson's condition is equivalent to the MFCQ when $\mathcal{X} = \mathbb{R}^N$.

*2) LICQ:* When $\mathcal{X} = \mathbb{R}^N$ and the rows of $\nabla \boldsymbol{h}(\boldsymbol{z}^*)$ as well as the gradients of the *active* inequality constraint functions $\nabla g_{i\ell}(\boldsymbol{x}_i^*)$'s are linearly independent, we can easily find $\boldsymbol{z}^{int}$ such that $\nabla \boldsymbol{h}(\boldsymbol{z}^*)(\boldsymbol{z}^{int} - \boldsymbol{z}^*) = \boldsymbol{0}$ and $\nabla g_{i\ell}(\boldsymbol{x}_i^*)(\boldsymbol{x}_i^{int} - \boldsymbol{x}_i^*) < 0$, $\forall \ell \in I_i(\boldsymbol{x}_i^*), \forall i$. This means that Robinson's condition is implied by the *linear independence constraint qualification* (LICQ).

*3) Slater's condition:* When the constraint set of problem $(P)$ is convex (i.e., $\boldsymbol{h}(\cdot)$ is affine and $\boldsymbol{g}_i(\cdot)$'s are convex) and the Slater's condition holds, i.e., there exists a point $\boldsymbol{z}^s = (\boldsymbol{x}^s, \boldsymbol{y}^s) \in int(\mathcal{X} \times \mathbb{R}^M)$ such that $\boldsymbol{h}(\boldsymbol{z}^s) = \boldsymbol{0}$ and $\boldsymbol{g}_i(\boldsymbol{x}_i^s) < \boldsymbol{0}$, $\forall i$, it can be easily shown that the following relations hold

$$\nabla g_{i\ell}(\boldsymbol{x}_i^*)(\boldsymbol{x}_i^s - \boldsymbol{x}_i^*) \leq g_{i\ell}(\boldsymbol{x}_i^s) - g_{i\ell}(\boldsymbol{x}_i^*) < 0, \; \forall \ell \in I_i(\boldsymbol{x}_i^*), \forall i,$$
$$\nabla \boldsymbol{h}(\boldsymbol{z}^*)(\boldsymbol{z}^s - \boldsymbol{z}^*) = \boldsymbol{h}(\boldsymbol{z}^s) - \boldsymbol{h}(\boldsymbol{z}^*) = \boldsymbol{0}.$$

Hence, the Slater's condition is sufficient for Robinson's condition for problems with convex constraints.

*4) Linearly constraint qualification:* When problem $(P)$ has linear constraints (i.e., $\mathcal{X} = \mathbb{R}^N$ and $\boldsymbol{h}(\cdot)$ and $\boldsymbol{g}_i(\cdot)$'s are affine), as in the case of Slater's condition, it can be readily verified that Robinson's condition holds true.

### B. Practical Considerations on Parameter Selection and Termination Conditions

In the PDD method, the control parameter $\eta_k$ determines how often the AL method and the penalty method are carried out. If $\eta_k$ is decreased too fast, then the penalty method will often take place, resulting in a large penalty and slow convergence. On the other hand, when $\eta_k$ is decrease very slowly, then the AL method will be more often performed. However, if the AL method does not converge in this case, such a choice will also slow down the convergence of the PDD. A more adaptive way to set $\eta_k$ is to make it explicitly related to the constraint violation. For example, we can set $\eta_k = \tau \min(\eta_{k-1}, \|\bm{h}(\bm{z}^{k-1})\|_\infty)$ where $0 < \tau < 1$. Similarly, the penalty parameter $\varrho_k$ can impact the convergence the PDD method. Specifically, when $\varrho_k$ decreases too fast, the AL problem will become ill-conditioned which impact the convergence of the optimization oracle. A simple way to set $\varrho_k$ is to let $\varrho_{k+1} = c\varrho_k$ where the parameter $c$ is a fraction which should be appropriately chosen to control the decreasing speed of the penalty parameter. Various choices of the parameter settings will be examined extensively in the second part of this paper.

Besides the parameter choice, the termination condition of the optimization oracle also affects the convergence of the PDD/IPDD. To guarantee theoretical convergence, we have used Eq. (18) to terminate the optimization oracle. However, it is sometime difficult to evaluate $\bm{e}^k$ and $\bm{\Delta}^k$ when the set $\mathcal{X}$ is complicated and the function $s_j(\bm{y}_j)$ is not simple. In practice, it is reasonable to terminate the optimization oracle based on the progress of the objective value $\mathcal{L}_k(\bm{z}^k)$, i.e.,

$$\frac{|\mathcal{L}_k(\bm{z}^k) - \mathcal{L}_{k-1}(\bm{z}^{k-1})|}{|\mathcal{L}_{k-1}(\bm{z}^{k-1})|} \le \epsilon_k.$$

Another practical choice of the termination condition for the optimization oracle is simple by setting the maximum number of iterations. Such termination condition is very suitable for in-network distributed implementation as it does not require coordination among network agents. Although the latter condition lacks theoretical guarantee, it is actually perform well in our numerical experience.

### C. Optimization oracle

To make use of the problem structure, we advocate using BCD-type algorithms as optimization oracle to address the AL problem $(P_{\varrho_k, \bm{\lambda}_k})$. Certainly, any other reasonable optimization method can be used, as long as they can guarantee the theoretical termination condition (18). For example, when some inequality constraint $\bm{g}_j(\bm{x}_j) \le 0$ is complicated, we can use concave-convex procedure [50], [51] to address the AL problem, i.e., we can replace $\bm{g}_j(\bm{x}_j)$ with its simple upper bound function [52] and solve the resulting problem (which is often easier) instead of the AL problem . In addition, when the AL problem can be globally solved by certain solver, the PDD/IPDD method with such global solver (instead of the BCD-type algorithms) could provide globally optimal solution to problem $(P)$.

## VI. CONCLUSIONS

In this paper, we design an optimization algorithm for a class of nonsmooth and nonconvex problems. The proposed algorithm, named PDD, can deal with difficult nonconvex coupling constraints, and it is further able to fully explore the problem structure for efficient numerical implementation. The PDD can be used to address a wide range of difficult engineering problems arising from areas such as signal processing, wireless communication and machine learning. In the second part of this paper we will demonstrate the strength of our algorithm by customizing it to a number of applications.

## APPENDIX A
## SOME BASICS

To improve the readability, we here list a few definitions and facts which are from [41, Chap 2&3] and used throughout the paper.

### A. Tangent cone, polar cone and normal cone

Tangent cone is the set of tangent directions whose definition is given as follows.

<u>Definition</u> A.1: [41, Def. 3.11] A direction $\bm{d}$ is called *tangent* to the set $X \subset \mathbb{R}^n$ at the point $\bm{x} \in X$ if there exist sequences of points $\bm{x}^k \in X$ and scalars $\tau_k > 0$, $k = 1, 2, \ldots$, such that $\tau_k \downarrow 0$ and

$$\bm{d} = \lim_{k \to \infty} \frac{\bm{x}^k - \bm{x}}{\tau_k}$$

Further, define the cone of feasible directions at $\bm{x} \in X$:

$$K_X(\bm{x}) = \{\bm{d} \in \mathbb{R}^n \mid \bm{d} = \beta(\bm{y} - \bm{x}), \bm{y} \in X, \beta \ge 0\}.$$

Then we have

<u>Lemma</u> A.1: [41, Lemma 3.13] Let $X \subset \mathbb{R}^n$ be a convex set and let $\bm{x} \in X$. Then the **tangent cone**, i.e., the set of tangent direction, of the set $X$ at $\bm{x}$ is

$$T_X(\bm{x}) = col\ K_X(\bm{x})$$

where $col\ X$ means the closure of the set $X$.

The polar cone is defined as follows.

<u>Definition</u> A.2: [41, Def. 2.23] Let $K$ be a cone in $\mathbb{R}^n$. The set

$$K^\circ \triangleq \{\bm{y} \in \mathbb{R}^n \mid \bm{y}^T \bm{x} \le 0\ \forall \bm{x} \in K\}$$

is called the **polar cone** of $K$.

Define

$$K \triangleq \{\bm{x} \in K_1 \mid \bm{A}\bm{x} \in K_2\}. \tag{51}$$

Given the definition of polar cone, the following fact holds true.

<u>Theorem</u> A.1: [41, Theorem 2.36] Assume that $K_1$ and $K_2$ are closed convex cones, and $K$ is defined by (51). If

$$0 \in \text{int}\{\bm{A}\bm{x} - \bm{y} : \bm{x} \in K_1, \bm{y} \in K_2\}, \tag{52}$$

then

$$K^\circ = K_1^\circ + \{\bm{A}^T \bm{\lambda} : \bm{\lambda} \in K_2^\circ\}. \tag{53}$$

The definition of normal cone is given as follows.

<u>Definition</u> A.3: [41, Def. 2.37] Let $X$ be a closed convex set and let $\bm{x} \in X$. Then

$$N_X(\bm{x}) = \{\bm{v} \in \mathbb{R}^n \mid \bm{v}^T(\bm{y} - \bm{x}) \le 0, \forall \bm{y} \in X\}$$

is called **normal cone** to $X$ at $\bm{x}$.

Following the above definitions, we have for a closed convex set $X$

$$[T_X(x)]^o = [K_X(x)]^o = N_X(x). \tag{54}$$

## B. Robinson's condition

Consider the problem

$$\begin{aligned}
\min\ & f(\boldsymbol{x}) \\
\text{s.t.}\ & g_i(\boldsymbol{x}) \leq 0, i = 1, \ldots, m, \\
& h_i(\boldsymbol{x}) = 0, i = 1, \ldots, p, \\
& \boldsymbol{x} \in X
\end{aligned} \quad (55)$$

with continuously differentiable functions $f : \mathbb{R}^n \to \mathbb{R}$, $\boldsymbol{g} : \mathbb{R}^n \to \mathbb{R}^m$, $\boldsymbol{h} : \mathbb{R}^n \to \mathbb{R}^p$ and with a closed convex set $X$. We consider a feasible point $\boldsymbol{x}_0$ of problem (55) and define the set of active inquality constraints:

$$I(\boldsymbol{x}_0) = \{1 \leq i \leq m : g_i(\boldsymbol{x}_0) = 0\}.$$

*Robinson's condition* with respect to the constraint set of problem (55) takes on the form

$$\left\{ \begin{pmatrix} \nabla \boldsymbol{h}(\boldsymbol{x}_0)\boldsymbol{d} \\ \nabla \boldsymbol{g}(\boldsymbol{x}_0)\boldsymbol{d} - \boldsymbol{v} \end{pmatrix} \middle| \boldsymbol{d} \in T_X(\boldsymbol{x}_0), \boldsymbol{v} \in \mathbb{R}^m, \right.$$
$$\left. v_i \leq 0, i \in I(\boldsymbol{x}_0) \right\} = \mathbb{R}^p \times \mathbb{R}^m. \quad (56)$$

Let $Z$ denote the feasible set of problem (55). Then we have

*Theorem A.2:* [41, Theorem 3.15] If Robinson's condition holds for problem (55) at $\boldsymbol{x}_0$, then $T_Z(\boldsymbol{x}_0)$ takes the form

$$T_Z(\boldsymbol{x}_0) = \{\boldsymbol{d} \in \mathbb{R}^n \mid \boldsymbol{d} \in T_X(\boldsymbol{x}_0), \nabla \boldsymbol{h}(\boldsymbol{x}_0)\boldsymbol{d} = \boldsymbol{0},$$
$$\nabla g_i(\boldsymbol{x}_0)^T \boldsymbol{d} \leq 0, i \in I(\boldsymbol{x}_0)\}. \quad (57)$$

## C. The boundedness of Lagrange multipliers

The following theorem gives a necessary optimality condition (i.e., KKT conation) for problem (55) with continuously differentiable function $f(\cdot)$.

*Theorem A.3:* [41, Theorem 3.25] Let $\hat{\boldsymbol{x}}$ be a local minimum of problem (55). Assume that at $\hat{\boldsymbol{x}}$ the constraint qualification condition[4] is satisfied for problem (55). Then there exist multipliers $\hat{\lambda}_i \geq 0, i = 1, \ldots, m$, and $\hat{\mu}_i \in \mathbb{R}, i = 1, \ldots, p$, such that

$$0 \in \nabla f(\hat{\boldsymbol{x}}) + \sum_{i=1}^m \hat{\lambda}_i \nabla g_i(\hat{\boldsymbol{x}}) + \sum_{i=1}^p \hat{\mu}_i \nabla h_i(\hat{\boldsymbol{x}}) + N_X(\hat{\boldsymbol{x}}), \quad (58)$$

and

$$\hat{\lambda}_i g_i(\hat{\boldsymbol{x}}) = 0, i = 1, \ldots, m. \quad (59)$$

Furthermore, it is shown in the following lemma that, once (58) and (59) are satisfied, the corresponding Lagrange multipliers are all bounded under Robinson's condition.

*Lemma A.2:* [41, Lemma 3.26] Let $\hat{\boldsymbol{x}}$ be a local minimum of problem (55) and let $\hat{\Lambda}(\hat{\boldsymbol{x}})$ be the set of Lagrange multipliers $\hat{\lambda} \in \mathbb{R}_+^m$ and $\hat{\mu} \in \mathbb{R}^p$ satisfying (58), (59).

1) The set $\hat{\Lambda}(\hat{\boldsymbol{x}})$ is convex and closed.
2) If problem (55) satisfies Robinson's condition at $\hat{\boldsymbol{x}}$, then the set $\hat{\Lambda}(\hat{\boldsymbol{x}})$ is also bounded.

---

[4]If any of the sufficient conditions for (57) is satisfied, we say that problem (55) satisfies the constraint qualification condition.

# Penalty Dual Decomposition Method For Nonsmooth Nonconvex Optimization—Part II: Applications

Qingjiang Shi, Mingyi Hong, Xiao Fu, Tsung-Hui Chang

*Abstract*—In Part I of this paper, we proposed and analyzed a novel algorithmic framework, termed penalty dual decomposition (PDD), for the minimization of a nonconvex nonsmooth objective function, subject to difficult coupling constraints. Part II of this paper is devoted to evaluation of the proposed methods in the following three applications, ranging from communication networks to data analytics: i) the max-min rate fair multicast beamforming problem; ii) the sum-rate maximization problem in multi-antenna relay broadcast networks; and iii) the volume-min based structured matrix factorization problem, which is often used in document topic modeling. By exploiting the structure of the aforementioned problems, we develop a new class of algorithms based on the PDD framework. Differently from the state-of-the-art algorithms, they are proven to achieve convergence to stationary solutions of the aforementioned nonconvex problems. Numerical results validate the efficacy of the proposed schemes.

*Index Terms*—Penalty dual decomposition, multicast beamforming, sum-rate maximization, matrix factorization.

## I. INTRODUCTION

In Part I of this paper, we have proposed a generic algorithm for optimizing the following nonconvex problem with coupling constraints:

$$\min_{\boldsymbol{x} \in \mathcal{X}, \boldsymbol{y}} \quad F(\boldsymbol{x}, \boldsymbol{y}) \triangleq f(\boldsymbol{x}, \boldsymbol{y}) + \sum_{j=1}^{n_y} \tilde{\phi}(\boldsymbol{y}_j) \quad \text{(P)}$$
$$\text{s.t.} \quad \boldsymbol{h}(\boldsymbol{x}, \boldsymbol{y}) = \boldsymbol{0},$$
$$\boldsymbol{g}_i(\boldsymbol{x}_i) \leq 0, \forall i$$

where $\boldsymbol{x} \triangleq (\boldsymbol{x}_1, \boldsymbol{x}_2, \ldots, \boldsymbol{x}_n)$ and $\boldsymbol{y} \triangleq (\boldsymbol{y}_1, \boldsymbol{x}_2, \ldots, \boldsymbol{x}_{n_y})$; $\tilde{\phi}(\boldsymbol{y}_j) = \phi_j(s_j(\boldsymbol{y}_j))$ is a composite function, with $s_j(\boldsymbol{y}_j)$ being a *convex* but possibly nondifferentiable function while $\phi_j(x)$ a *nondecreasing* and *continuously differentiable* function; the feasible set $\mathcal{X}$ is the Cartesian product of $n$ simple *closed convex* sets $\mathcal{X}_i$'s with $\boldsymbol{x}_i \in \mathcal{X}_i$, $\forall i$; $f(\boldsymbol{x}, \boldsymbol{y})$ and each component of the vector functions $\boldsymbol{h}(\boldsymbol{x}, \boldsymbol{y})$ and $\boldsymbol{g}_i(\boldsymbol{x}_i)$'s are all continuously differentiable functions.

Our proposed algorithm, termed penalty dual decomposition (PDD), is a combination of primal dual based augmented Lagrangian method, block-coordinate-descent-type algorithm, and the penalty method. Under certain constraint qualification (CQ) named Robinson's condition, we show that every limit point generated by the PDD method is a KKT solution of problem (P). As will be shown in Part II of the paper, the main advantage of the PDD method is that, it is capable of exploiting the problem structure in a way which results in computationally lightweight algorithms. Particularly, by introducing appropriate auxiliary variables, the subproblem involved in the PDD method can be efficiently solved. As a result, the PDD method could be very efficient in dealing with nonconvex problems with difficult coupling constraints.

Many engineering problems can be formulated as problem (P). Some examples include dictionary learning and compressive sensing [2]–[4], volume-minization based matrix factorization [5]–[7], joint transceiver optimization of wireless systems [8]–[14], waveform design for radar systems [15]–[17], cross-layer design of wireless networks [18]–[21], and sensor network localization [22]–[24], etc.. Among the potential applications stated above, Part II of this paper focuses on performing case-studies on the following three important signal processing applications:

1) **Maxi-Min Fair Multicast beamforming**. Multicast beamforming is an important component of the evolved multimedia broadcast multicast service (eMBMS) in the long-term evolution (LTE) standard [25]. In multicast beamforming, a base station (BS) with multi-antennas transmits common information to multiple groups of users. For efficient multicasting, the BS chooses different weights for different streams based on the channel state information (CSI) to steer the transmit power in the directions of each group of users while limiting inter-group interference. To guarantee the rate fairness among the users, we often design bemforming weights to maximize the minimum user rate subject to a BS power constraint. This problem is known to be NP-hard [10] and has received a lot of attention from the research community [10], [12], [13]. A well-known approach to dealing with the multicast beamforming problem is using the celebrated semidefinite relaxation (SDR) method [10], [12]. However, to recover a high-quality suboptimal solution, *Gaussian randomization procedure* is needed after solving the SDR problem, resulting in high computational complexity. In this paper, we propose a PDD-based iterative multicast beamforming algorithm which achieves a higher max-min user rate but lower complexity than the SDR method.

2) **Sum-rate maximization for relay broadcast channel**. Relay-based cooperative communication has been adopted in LTE-Advanced standard as a key technology for future generation wireless communication systems [26]. In a relay-assisted cellular downlink system, the link quality between the BS and cell-edge users would benefit from deploying a relay as well as joint source (i.e., BS)-relay design. However, the relay transmission introduces a coupling between the source precoder and the relay precoder in relay power constraints, which poses a fundamental challenge in joint source-relay design. In fact, such a challenge exists in various relay-assisted communication systems, e.g., multi-hop relay networks [27], [28], two-way relay networks [29], [30], relay interference networks [31], [32], etc.. Despite having extensive research on relay systems, there is still a lack of efficient optimization method to address the difficulty arising from the intrinsic coupling between the source precoder and the relay precoder. A promising way to address the coupling of two precoders [31], [33]–[35] is using alternating optimization (AO), i.e., alternatingly optimizes one precoder while fixing the other. However, the AO scheme can easily get trapped in some non-stationary solutions, which can have very low system throughput. In this paper, by applying PDD, we propose an efficient optimization framework to deal with the coupling between the source precoder and the relay precoder in joint design of relay systems.

3) **Volume-min based matrix factorization**. Structured factorization for given data matrices has many applications in signal processing and machine learning [36], [37]. As one important criterion for structured matrix factorization, volume minimization (VolMin) finds the minimum-volume *simplex* that embraces all the given data points [36]. This criterion can guarantee the identifiability of the factor matrices under mild conditions that are realistic in a wide variety of applications [7]. Hence, it recently attracted considerable interest in document clustering [5], blind separation of power spectral for dynamic spectrum access [38], and remote sensing [6], etc.. Due to the nature of matrix factorization, there exists a coupling of two matrix factors in the VolMin problem, making the VolMin problem quite challenging. In the literature, the VolMin problem is first transformed into the dimension-reduced space and then solved using alternating optimization or penalty method [6]. However, the existing algorithms cannot guarantee stationary solution to the VolMin problem. Moreover, in applications with additional constraints (e.g., nonnegativity of matrix factors), the problem has to be solved in the *original* space, rather than in the dimension-reduced space where the existing algorithms do no work. In this paper, we propose a PDD-based VolMin algorithm which works well in the original data space with guaranteed convergence.

The key to applying the PDD framework to nonconvex problems is to properly reformulate the problems at hand as problem $(P)$, so that the corresponding augmented Lagrange problems can be easily solved via block-coordinate-descent (BCD)-type algorithms. In this paper, we present some reformulations of the aforementioned three problems in the form of problem $(P)$ and develop a new class of algorithms for the reformulations by applying the PDD framework. By applying BCD-type algorithms to these reformulations, we can fully exploit the problem structures of the aforementioned problems, and significantly alleviate the challenging nonconvexity arising from either objective functions (e.g., the max-min structure in the max-min fair multicasting beamforming problem, and the volume function in the VolMin problem) or the constraints (e.g., the relay power constraint in the joint source-relay design problem that couples the source precoder and the relay preocder, and the matrix factorization equality constraint in the VolMin problem).

The developed algorithms enjoy several desirable features. First, differently from the state-of-the-art algorithms, they are proven to achieve convergence to stationary solutions of the aforementioned problems, and in practical they outperform the state-of-the-art algorithms in a number of performance metrics. Second, the iterations of the algorithms have closed-form and simple updates, therefore they are relatively easy to implement. Third, the algorithms are quite flexible and they are applicable to some generalizations of the aforementioned problems as well. For instance, the PDD algorithm can be easily generalized to dealing with the VolMin problem in the original space with a nonnegativity constraint on the basis factors.

The remainder of this paper is organized as follows. In Section II-IV, we apply the PDD method to the the aforementioned problems. Specifically, in each of three sections, we first reformulate the three problems, then show how the corresponding augmented Lagrangian problem is solved by BCD-type algorithms, followed by some simulations to compare the performance of the PDD-based algorithms with the state-of-art algorithms. Section V concludes the paper.

*Notations*: Besides the notations specified in Part I of this paper, we use the following notations. $\mathbb{C}^n$ (or $\mathbb{C}^{m \times n}$) denotes the $n$ (or $m \times n$)-dimensional space of complex number. For a matrix $\mathbf{X}$, $\mathbf{X}^H$ and $\sigma_i(\mathbf{X})$ denote its conjugate transpose and its $i$-th largest singular value, respectively. For a vector $\boldsymbol{x}$, $\mathrm{diag}\{\boldsymbol{x}\}$ denotes a diagonal matrix with the elements of $\boldsymbol{x}$ being its diagonal entries. $\Re e\{x\}$ and $\Im m\{x\}$ denote the real part and the imaginary part of a complex number $x$, respectively, and $x^*$ denotes the conjugate of $x$. The notation $\mathbf{A} \otimes \mathbf{B}$ means the Kronecker product of two matrices $\mathbf{A}$ and $\mathbf{B}$. $\mathbf{A} \succeq 0$ (or $\succ 0$) means that $\mathbf{A}$ is a positive semidefinite (or definite) matrix. $\mathbb{E}\{\cdot\}$ denotes expectation operation.

## II. MAXMIN-RATE FAIRNESS MULTI-CAST BEAMFORMING

Signal-to-interference-plus-noise ratio is an important performance metric used in signal design. It is generally in a *quadratic ratio* form with respect to the designed variables. On the other hand, max-min fairness is a popular resource allocation criterion that is widely adopted in wireless communication and signal processing [10], [12], [13], [39], [40]. As a result, we are often faced with the following problem

$$\max_{\boldsymbol{x} \in \mathcal{X}} \min_{k \in \mathcal{K}} \frac{\boldsymbol{x}^H \mathbf{A}_k \boldsymbol{x}}{\boldsymbol{x}^H \mathbf{B}_k \boldsymbol{x}} \qquad (1)$$

where $\boldsymbol{x}$ is a design variable which is constrained to a set $\mathcal{X}$; $\mathbf{A}_k$'s and $\mathbf{B}_k$'s are known matrices with $\mathbf{B}_k \succ 0$. Several examples of (1) can be found in max-min fariness precoding for wireless networks [14], [41], waveform design for radar systems [39], [40], and robust classification in machine learning [42], etc.. Problem (1) is challenging due to the nonlinear and nondifferentiable max-min ratio structure. In this section, as an important example, we illustrate the application of PDD to multi-cast beamforming for achieving max-min rate fairness [10].

### A. Problem Formulation

Consider a single-cell multi-user multiple-input-single-output (MISO) downlink system, where a base station (BS) equipped with $N_t$ antennas transmits $n_g > 1$ independent data streams to $n_g$ group of users over a common frequency band. Suppose that the $i$-th group, denoted by $\mathcal{G}_i$, has $m_i$ single-antenna users, each of which is interested in receiving a common data stream. Let $s_i$ denote the data stream for group $\mathcal{G}_i, i = 1, 2, \ldots, n_g$ and $\boldsymbol{w}_i \in \mathbb{C}^{N_t}$ be the beamforming weight for the $i$-th group. The transmitted signal at the BS is given by $\sum_{i=1}^{n_g} \boldsymbol{w}_i s_i$. Let $\boldsymbol{h}_k \in \mathbb{C}^{N_t}$ denote the conjugated channel between the BS and the receiver $k \in \mathcal{G}_i$. Then the received signal at receiver $k \in \mathcal{G}_i$ is given by

$$r_k = \boldsymbol{h}_k^H \boldsymbol{w}_i s_i + \sum_{j \neq i} \boldsymbol{h}_k^H \boldsymbol{w}_j s_j + z_k, \quad k \in \mathcal{G}_i \quad (2)$$

where $z_k$ denotes additional Gaussian white noise (AWGN) with variance $\sigma_k^2$.

Assume that $s_i$'s are i.i.d complex Gaussian random variable with zero mean and unit variance, and moreover $s_i$'s and $z_k$'s are independent of each other. Then the signal-to-interference-plus-noise-ratio (SINR) can be expressed as

$$\text{SINR}_k = \frac{\boldsymbol{w}_i^H \mathbf{R}_k \boldsymbol{w}_i}{\sum_{j \neq i} \boldsymbol{w}_j^H \mathbf{R}_k \boldsymbol{w}_j + \sigma_k^2}, k \in \mathcal{G}_i, i = 1, 2, \ldots, n_g \quad (3)$$

where $\mathbf{R}_k \triangleq \boldsymbol{h}_k \boldsymbol{h}_k^H$.

To achieve rate fairness among users, a popular criterion for beamforming design is to maximize the minimum user rate subject to the BS power constraint $\sum_{i=1}^{n} \|\boldsymbol{w}_i\|^2 \leq P_{BS}$, where $P_{BS}$ denotes the total available power at the BS. Since the power constraint must be active at the optimality, we can write the max-min rate fairness multi-cast beamforming problem equivalently as

$$\max_{\{\boldsymbol{w}_i\}} \min_i \min_{k \in \mathcal{G}_i} \quad \log_2\left(1 + \frac{\boldsymbol{w}^H \mathbf{A}_{i_k} \boldsymbol{w}}{\boldsymbol{w}^H \mathbf{B}_{i_k} \boldsymbol{w}}\right), \quad (4)$$
$$\text{s.t.} \quad \|\boldsymbol{w}\|^2 = 1$$

where $\boldsymbol{w} = (\boldsymbol{w}_i)_i$, $\mathbf{A}_{i_k} = \text{diag}\{\boldsymbol{e}_i\} \otimes \mathbf{R}_k$, and

$$\mathbf{B}_{i_k} = (\mathbf{I} - \text{diag}\{\boldsymbol{e}_i\}) \otimes \mathbf{R}_k + \frac{\sigma_k^2}{P_{BS}} \mathbf{I}.$$

This problem is known to be NP-hard [10]. After solving (4), we need to scale $\boldsymbol{w}$ such the power constraint. A popular method to address this problem is using semidefinite relaxation method coupled with bisection method [10], referred to as BisecSDR method, where, in each bisection, it is required to solve a semidefinite programming, requiring complexity at most $O\left(I_{bsc} \log(\frac{1}{\epsilon_{ip}}) \sqrt{n_g N_t} (n_g^3 N_t^6 + n_g N_t^2 K)\right)$. Here $K \triangleq \sum_{i=1}^{n_g} m_i$, the parameter $\epsilon_{ip}$ represents the solution accuracy at the interior-point algorithm's termination, and $I_{bsc}$ denotes the number of bisections.

### B. PDD-based Algorithm

For convenience, let us consider a more general but equivalent formulation of problem (4), which is given by

$$\boxed{\max_{\boldsymbol{w}} \min_{k \in \mathcal{K}} \quad \frac{\boldsymbol{w}^H \mathbf{A}_k \boldsymbol{w}}{\boldsymbol{w}^H \mathbf{B}_k \boldsymbol{w}}, \quad \text{s.t.} \quad \|\boldsymbol{w}\|^2 = 1} \quad (P1)$$

where $\mathcal{K} \triangleq \{1, 2, \ldots, K\}$, and the matrices $\mathbf{A}_k$'s are all positive semidefinite and $\mathbf{B}_k$'s are all positive definite. In what follows, we present the PDD-based algorithm for problem $(P1)$.

First, we recast problem $(P1)$ as follows

$$\max_{\boldsymbol{t} \geq 0, \boldsymbol{w}} \quad \min_k t_k$$
$$\text{s.t.} \quad \|\mathbf{A}_k^{\frac{1}{2}} \boldsymbol{w}\| = t_k \|\mathbf{B}_k^{\frac{1}{2}} \boldsymbol{w}\|, \quad \forall k, \quad (5)$$
$$\|\boldsymbol{w}\|^2 = 1,$$

which is a special case of problem $(P)$. In problem (5), the first $K$ equality constraints are difficult coupling constraints. By moving these constraints into the objective, we obtain the corresponding augmented Lagrangian problem as follows

$$\max_{\boldsymbol{t} \geq 0, \boldsymbol{w}} \quad \min_k t_k - \frac{1}{2\rho} \sum_{k=1}^{K} \left(\|\mathbf{A}_k^{\frac{1}{2}} \boldsymbol{w}\| - t_k \|\mathbf{B}_k^{\frac{1}{2}} \boldsymbol{w}\| + \rho \lambda_k\right)^2$$
$$\text{s.t.} \quad \|\boldsymbol{w}\|^2 = 1. \quad (6)$$

where $\rho$ is a penalty parameter and $\lambda_k$ is a Lagrange multiplier associated with the $k$-th constraint.

The key to using the PDD method is to find appropriate locally tight lower bounds for the objective function, so that BSUM [43] can be applied to optimize the AL. For problem (6), we can simply decouple the variables into two blocks $\boldsymbol{w}$ and $\boldsymbol{t}$, leading to two subproblems: i.e., 1) solve (6) for $\boldsymbol{t}$ while fixing $\boldsymbol{w}$, and 2) solve (6) for $\boldsymbol{w}$ while fixing $\boldsymbol{t}$, which are respectively referred to as *$\boldsymbol{t}$-subproblem* and *$\boldsymbol{w}$-subproblem*. The $\boldsymbol{t}$-subproblem is strictly convex and thus has a unique solution, which can be easily solved by exploiting the problem structure; see Appendix A for a detailed derivation. The main difficulty lies in solving the $\boldsymbol{w}$-subproblem given by

$$\min_{\boldsymbol{w}} \quad \vartheta(\boldsymbol{w}) \triangleq \sum_{k=1}^{K} \left(\|\mathbf{A}_k^{\frac{1}{2}} \boldsymbol{w}\| - t_k \|\mathbf{B}_k^{\frac{1}{2}} \boldsymbol{w}\| + \rho \lambda_k\right)^2$$
$$\text{s.t.} \quad \|\boldsymbol{w}\|^2 = 1.$$

Apparently, the $\boldsymbol{w}$-subproblem is difficult to solve. Instead of exactly minimizing $\vartheta(\boldsymbol{w})$, we try to find a locally tight upper bound $u(\boldsymbol{w}; \tilde{\boldsymbol{w}})$ for $\vartheta(\boldsymbol{w})$ and minimize this upper bound to update $\boldsymbol{w}$ given $\boldsymbol{t}$. Observing the constraint $\|\boldsymbol{w}\| = 1$, we expect the upper bound to be a *homogeneous quadratic function* in the form of $\boldsymbol{w}^H \mathbf{C} \boldsymbol{w}$ or $\boldsymbol{w}_{eq}^T \mathbf{C} \boldsymbol{w}_{eq}$ where $\boldsymbol{w}_{eq} \triangleq (\Re e\{\boldsymbol{w}\}, \Im m\{\boldsymbol{w}\})$, so that the resulting problem is an easily solvable eigenvalue problem.

By expanding $\vartheta(\boldsymbol{w})$, we can find that $\vartheta(\boldsymbol{w})$ includes the following four kinds of terms: 1) $\boldsymbol{w}^H \mathbf{A}_k \boldsymbol{w} + t_k^2 \boldsymbol{w}^H \mathbf{B}_k \boldsymbol{w}$; 2) $-2t_k \|\mathbf{A}_k^{\frac{1}{2}} \boldsymbol{w}\| \|\mathbf{B}_k^{\frac{1}{2}} \boldsymbol{w}\|$; 3) $2\rho \lambda_k \|\mathbf{A}_k^{\frac{1}{2}} \boldsymbol{w}\|$; 4) $-2\rho \lambda_k t_k \|\mathbf{B}_k^{\frac{1}{2}} \boldsymbol{w}\|$. Clearly, we need to make efforts to bound the last three terms with homogenous quadratic functions. Unfortunately, since the multiplier $\lambda_k$'s could be either negative or positive, it is challenging to bound the last two terms with homogenous quadratic functions. Thanks to the fact that $\|w\|=1$, we can modify the third term as $2\rho\lambda_k\|\mathbf{A}_k^{\frac{1}{2}}w\|\|w\|$ when $\lambda_k < 0$ and the fourth term as $-2\rho\lambda_k t_k\|\mathbf{B}_k^{\frac{1}{2}}w\|\|w\|$ when $\lambda_k > 0$. Hence, essentially, $\vartheta(w)$ includes two kinds of terms in the forms of $\|\mathbf{Q}_1 w\|$ and $-\|\mathbf{Q}_1 w\|\|\mathbf{Q}_2 w\|$ with some appropriate $\mathbf{Q}_1$ and $\mathbf{Q}_2$. To bound these two terms, we resort to the following lemma.

TABLE I
ALGORITHM 1: BSUM FOR PROBLEM (6)

---
0. initialize $w$ and $t$
1. **repeat**
2. $\quad a_k \longleftarrow \frac{\|\mathbf{B}_k^{\frac{1}{2}}w\|^2}{2\rho}$
3. $\quad b_k \longleftarrow \frac{\|\mathbf{A}_k^{\frac{1}{2}}w\| + \rho\lambda_k}{\|\mathbf{B}_k^{\frac{1}{2}}w\|}$.
4. $\quad$ update $t$ by solving problem (48)
5. $\quad$ compute $\mathbf{C} = \sum_{k=1}^K \mathbf{C}_k$ via (57) with $\tilde{w} = w$
6. $\quad$ update $w \longleftarrow v_{\min}(\mathbf{C})$
7. **until** some termination criterion is met
---

*Lemma 2.1:* For real vectors $x$, $y$, $\tilde{x}$, $\tilde{y}$, the following inequalities
1) $\|x\|\|y\| \geq \frac{1}{\|\tilde{x}\|\|\tilde{y}\|} x^T \tilde{x}\tilde{y}^T y$, $\forall \tilde{x} \neq 0, \tilde{y} \neq 0, x, y$;
2) $\|x\| \leq \frac{1}{2\|\tilde{x}\|}\|x\|^2 + \frac{1}{2}\|\tilde{x}\|$, $\forall \tilde{x} \neq 0, x$

hold true with equality satisfied at $x = \tilde{x}$ and $y = \tilde{y}$.

*Proof:* Part 1) follows directly from the Cauchy-Schwartz inequality, while Part 2) follows from the property of concave function by noting that $\|x\| = \sqrt{\|x\|^2}$ is a concave function of $\|x\|^2$. ∎

In terms of the above analysis and using Lemma 2.1, we can obtain $\vartheta(w) \leq u(w, \tilde{w}) \triangleq w_{eq}^T \mathbf{C} w_{eq} + const$ in the real domain, where $\mathbf{C}$ is a $2n_g N_t$ by $2n_g N_t$ matrix function of $\tilde{w}$ whose detailed derivation is shown in Appendix B. Moreover, it can be verified that $u(w, \tilde{w})$ is a locally tight upper bound [43] of $\vartheta(w)$ over the set $\{w \mid \|w\|=1\}$. With such an upper bound function, we update $w$ by solving the following eigenvalue problem, i.e., $\min_{w_{eq}} w_{eq}^T \mathbf{C} w_{eq}, \text{s.t.} \|w_{eq}\| = 1$. Denote by $v_{min}(\mathbf{C})$ the eigenvector of $\mathbf{C}$ corresponding to its minimum eigenvalue. Once we get $v_{min}(\mathbf{C})$, we can construct the corresponding $w$.

To summarize, the BSUM algorithm for addressing problem (6) is presented in TABLE I. It can be shown that the most costly step of the BSUM algorithm lies in calculating $v_{min}(\mathbf{C})$, requiring complexity of $O(Kn_g^2 N_t^2) + O(n_g^3 N_t^3)$, where the first term corresponds to the computation of $\mathbf{C}$ while the second term corresponds to the eigenvalue decomposition. It is easily seen that the PDD method has lower complexity than the BisecSDR method in [10].

### C. Numerical Results

In the simulations, the noise power is set to unit for all receivers and $P_{BS} = 10$ dB. For convenience, we denote by $(N_t, n_g, m_g)$ a multi-user multi-cast network with $N_t$ BS antennas, $n_g$ multi-cast groups each with $m_g$ single-antenna users, hence $K = n_g m_g$ users in total. Furthermore, unless otherwise specified, we set $\rho_0 = 0.5K$, $\epsilon_0 = 1e-3$, $\epsilon_O = 1e-4$, as well as $\rho_k = c\rho_{k-1}$ and $\epsilon_k = \epsilon_{k-1}c$ with $c = 0.6$ for the PDD method in all of our simulations. Moreover, to avoid numerical instability, we set the maximum number of inner BSUM iterations of the PDD method as 100 in practical implementation.

We compare the PDD method with the BisecSDR method in [10] and the penalty-BSUM method[1] proposed in [45] (abbreviated as "Penalty" in the plot). The basic idea of the BisecSDR method is as follows. First, by applying semidefinite relaxation (SDR), problem $(P1)$ is relaxed as

$$\max_{t,\mathbf{W}} \quad t$$
$$\text{s.t.} \quad \text{Tr}\left((\mathbf{A}_k - t\mathbf{B}_k)\mathbf{W}\right) \geq 0, \forall k, \quad (7)$$
$$\mathbf{W} \succeq 0.$$

Second, by searching over $t$ using Bisection method, we can obtain the optimal $\mathbf{W}$ by solving a sequence of semidefinite programmings (obtained by fixing $t$ in the above problem). Third, given the optimal $\mathbf{W}$, we can find a suboptimal solution $w$ by checking all candidate solutions including the principal eigenvector of $\mathbf{W}$ and those obtained by performing Gaussian randomization procedure (GRP). Note that, the optimal value of problem (7) can serve as an upper bound for the achievable maxmin user rate. Particularly, when the SDR is tight, the maxmin user rate coincides with the upper bound. In simulations, the semidefinite programmings are solved by interior-point method, e.g., using the off-the-shelf package SeDuMi [46] for efficiency. The Bisection procedure is terminated when the relative size of the bisection interval is smaller than $1e-3$. In addition, the penalty-BSUM shares the same parameter setting with the PDD.

The average convergence behavior of the algorithm over ten randomly generated examples is illustrated in Fig. 1, where the minimum user rate is normalized by the upper bound value. It is seen that the PDD method exhibits better convergence behavior than the penalty-BSUM method in terms of both the objective value and the optimality gap, while both achieving similar constraint violation. Here the optimality gap measures how well the solution $w$ satisfies the KKT condition of problem $(P1)$, which is defined by the optimal value of the following convex optimization problem[2]

$$\min_{\{\lambda_k\}} \quad \left\|\sum_{k=1}^K \lambda_k \boldsymbol{f}_k + \lambda_0 w\right\|$$
$$\text{s.t.} \quad \sum_{k=1}^K \lambda_k = 1, \lambda_k \geq 0, k = 1, 2, \ldots, K \quad (8)$$

---
[1]The penalty-BSUM algorithm is similar to the PDD method but does not include the dual update as in the PDD method. The algorithm in [44] is in essence the penalty-BSUM algorithm, with the only difference in that some fixed penalty parameter was used in [44] while the penalty-BSUM algorithm uses increasing penalty. However, fixed penalty parameter cannot guarantee a KKT solution. Moreover, it is generally difficult to choose a penalty parameter which works well for all cases. Hence, we modify the algorithm in [44] to the exact penalty-BSUM algorithm by using increasing penalty.

[2]By KKT analysis, it can be shown that problem (8) having a zero optimal value is a necessary optimality condition for problem $(P1)$.

where $\boldsymbol{f}_k$ is the gradient of the function $\frac{\boldsymbol{w}^H \mathbf{A}_k \boldsymbol{w}}{\boldsymbol{w}^H \mathbf{B}_k \boldsymbol{w}}$ with respect to $\boldsymbol{w}$. Moreover, the PDD can achieve the upper bound value in this example, implying the excellent performance of the PDD method. In addition, one can see that both the feasibility gap and the optimality gap (i.e., constraint violation defined in Part I) decrease at the same time. Although the zero feasibility gap does not necessarily imply the zero optimality gap, it is much easier to evaluate the former than the latter. Hence, in our later simulations, we only examine the feasibility gap for simplicity.

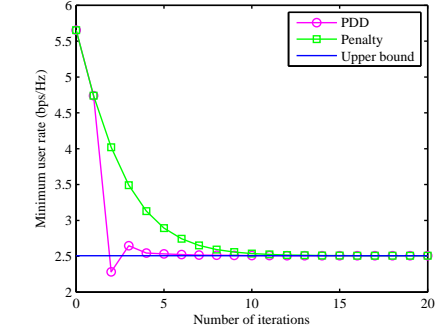

(a) The minimum user rate.

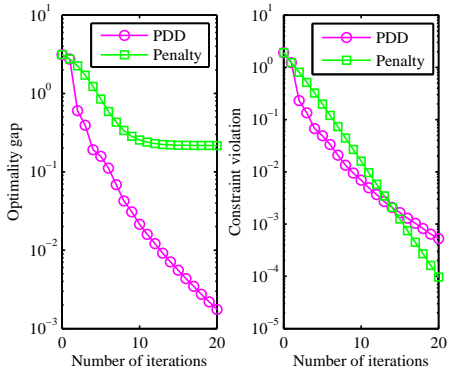

(b) The optimality and feasibility gap.

Fig. 1. The convergence behavior of the PDD method for network $(8, 4, 2)$.

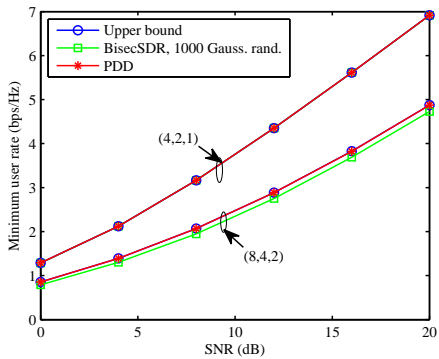

Fig. 2. The average minimum user rate achieved by various methods versus SNR for different networks.

Fig. 2 illustrates the max-min rate performance of the PDD method versus the BS power budget $P_{BS}$ as compared to the upper bound provided by the BisecSDR method (i.e., the optimal value of problem (7)) and the performance of the BisecSDR method with 1000 Gaussian randomizations. The results in the plot are averaged over 100 random channel realizations. For the network $(4, 2, 1)$, it is known that the SDR is tight in this case and thus the upper bound is exactly the same as the optimal max-min user rate. From the figure, it is observed that the performance of the PDD method coincides with the upper bound for both networks, which is better than the performance of the BisecSDR method for the network $(8, 4, 2)$.

For a clear illustration, Table 2 compares the performance of three methods in terms of the cpu time and the achieved minimum rate averaged over 100 random channel realizations. In the table, $R_{UB}$, $R_{PDD}$, $R_{SDR}$, and $R_{Penalty}$ denote the upper bound value, and the minimum rate achieved by the PDD method, the BisecSDR method with 1000 Gaussian randomizations, and the penalty-BSUM method, respectively, while $T_{PDD}$, $T_{SDR}$, and $T_{Penalty}$ denote the corresponding cpu time required by three methods. It can be observed that the PDD method requires less cpu time than the BisecSDR method while achieving almost global optimality. Moreover, it performs more efficiently than the penalty-BSUM method in terms of the consumed cpu time.

TABLE II
THE AVERAGE CPU TIME AND MIN. RATE COMPARISON

| Network | $\frac{R_{PDD}}{R_{UB}}$ | $\frac{R_{PDD}}{R_{SDR}}$ | $\frac{T_{SDR}}{T_{PDD}}$ | $\frac{R_{PDD}}{R_{Penalty}}$ | $\frac{T_{Penalty}}{T_{PDD}}$ |
|---|---|---|---|---|---|
| $(2, 2, 2)$ | 99.97% | 100.36% | 3.38 | 100.26% | 2.05 |
| $(4, 2, 2)$ | 99.98% | 100.28% | 3.83 | 100.14% | 1.98 |
| $(8, 4, 2)$ | 99.98% | 102.06% | 3.44 | 100.13% | 1.96 |
| $(8, 2, 4)$ | 99.93% | 101.78% | 3.02 | 100.15% | 1.74 |
| $(16, 4, 4)$ | 99.92% | 102.93% | 2.76 | 100.18% | 1.70 |

## III. JOINT SOURCE-RELAY DESIGN FOR MULTI-ANTENNA RELAY BROADCAST SYSTEMS

Wireless relaying in cellular networks has attracted considerable attention due to its advantage of coverage extension and throughput improvement. It is well-known that, joint source-relay design can further enhance the system throughput performance for multi-antenna relay systems. However, the relay power constraint results in the coupling between the source precoder and the relay precoder, therefore the resulting joint source-relay design problem is very challenging to solve. In this section, by applying PDD, we present a joint source-relay design method which can reach at least stationary solutions. Note that our method is developed for a multi-antenna relay broadcasting channel but its basic idea can be extended to joint source-relay design of other relay systems.

### A. Problem formulation

Consider a sum-rate maximization problem for a multi-antenna relay broadcasting channel, where a multi-antenna source (e.g, base station), equipped with $N_s$ antennas, sends signal to $K$ single-antenna users with the aid of a multi-antenna relay equipped with $N_r$ antennas. The received signal

at each user can be expressed as

$$\boldsymbol{y}_k = \boldsymbol{g}_k^H \mathbf{F} \left( \mathbf{H} \sum_{j=1}^{K} \boldsymbol{v}_j s_j + \boldsymbol{n}_R \right) + n_k, \quad k=1,2,\ldots,K. \quad (9)$$

where $\boldsymbol{v}_j$ and $s_j$ denote transmit beamformer employed by the source and the transmitted symbol intended for user $j$, respectively; the term $\sum_{j=1}^{K} \boldsymbol{v}_j s_j$ is the transmit signal of the source; $\mathbf{H} \in \mathbb{C}^{N_r \times N_s}$ represents the channel between the source and the relay; $\boldsymbol{n}_R$ and $n_k$ denote the AWGN at the relay and user $k$, respectively; $\mathbf{F} \in \mathbb{C}^{N_r \times N_r}$ is the precoder employed by the relay to process the received signal (i.e., the bracketed term) from the source; $\boldsymbol{g}_k \in \mathbb{C}^{N_r}$ denotes the conjugated channel between the relay and user $k$.

Suppose that the transmitted symbols and noises are independent of each other. Moreover, let $\sigma_R^2$ and $\sigma_k^2$ denote the noise power at the relay and user $k$, and define the source precoder $\mathbf{V} \triangleq [\boldsymbol{v}_1 \ \boldsymbol{v}_2 \ \ldots \ \boldsymbol{v}_K] \in \mathbb{C}^{N_s \times K}$. Then the SINR $\gamma_k$ at user $k$ is given by

$$\gamma_k(\mathbf{V}, \mathbf{F}) \triangleq \frac{|\boldsymbol{g}_k^H \mathbf{F} \mathbf{H} \boldsymbol{v}_k|^2}{\sum_{j \neq k} |\boldsymbol{g}_k^H \mathbf{F} \mathbf{H} \boldsymbol{v}_j|^2 + \sigma_R^2 \|\boldsymbol{g}_k^H \mathbf{F}\|^2 + \sigma_k^2}. \quad (10)$$

Furthermore, the source power consumption is given by $\mathrm{Tr}(\mathbf{VV}^H)$ and the relay power consumption is $\|\mathbf{FHV}\|_F^2 + \sigma_R^2 \|\mathbf{F}\|_F^2$.

We are interested in maximizing the weighted sum-rate subject to the source and relay power constraints, which can be mathematically formulated as follows

$$\boxed{\begin{aligned} \max_{\mathbf{V},\mathbf{F}} & \sum_{k=1}^{K} \alpha_k \log\left(1 + \gamma_k(\mathbf{V},\mathbf{F})\right) \\ \text{s.t.} \ & \mathrm{Tr}(\mathbf{VV}^H) \leq P_S, \\ & \|\mathbf{FHV}\|_F^2 + \sigma_R^2 \|\mathbf{F}\|_F^2 \leq P_R. \end{aligned}} \quad (P2)$$

where $\alpha_k$ denotes the weight measuring the priority of user $k$, $P_S$ and $P_R$ denote the source and relay power budget. The problem is hard to solve due mainly to the coupling of the source precoder $\mathbf{V}$ and relay precoder $\mathbf{F}$ at the relay power constraint. Note that such coupling is common to a number of joint source-relay designs well beyond problem (P2). We here aim to provide a way to deal with such coupling constraints.

### B. PDD-based algorithm

We start by reformulating problem (P2) so that the PDD algorithm can be easily applied. Introducing a set of auxiliary variables $\{\mathbf{X}, \bar{\mathbf{V}}, \bar{F}, \bar{\mathbf{X}}\}$, and defining $\mathcal{X} \triangleq \{\mathbf{V}, \mathbf{F}, \mathbf{X}, \bar{\mathbf{V}}, \bar{F}, \bar{\mathbf{X}}\}$ for notational simplicity, we can recast problem (P2) as

$$\max_{\mathcal{X}} \sum_{k=1}^{K} \alpha_k \log\left(1 + \frac{|\boldsymbol{g}_k^H \boldsymbol{x}_k|^2}{\sum_{j \neq k} |\boldsymbol{g}_k^H \boldsymbol{x}_j|^2 + \sigma_R^2 \|\boldsymbol{g}_k^H \mathbf{F}\|^2 + \sigma_k^2}\right)$$

$$\text{s.t.} \ \mathrm{Tr}(\bar{\mathbf{V}}\bar{\mathbf{V}}^H) \leq P_S,$$
$$\|\bar{\mathbf{X}}\|_F^2 + \sigma_R^2 \|\bar{\mathbf{F}}\|_F^2 \leq P_R, \quad (11)$$
$$\mathbf{X} = \mathbf{FHV}, \ \sigma_R \mathbf{F} = \sigma_R \bar{\mathbf{F}},$$
$$\mathbf{X} = \bar{\mathbf{X}}, \ \mathbf{V} = \bar{\mathbf{V}}.$$

where having $\sigma_R$ in the constraint $\sigma_R \mathbf{F} = \sigma_R \bar{\mathbf{F}}$ facilitates the solution of the subproblem involving $\{\bar{\mathbf{X}}, \bar{\mathbf{F}}\}$. This point will become clear shortly. Now we can see that the reformulation has separable inequality constraints. By building all the equality constraints into the objective, we can obtain the augmented Lagrangian problem as follows

$$\max_{\mathcal{X}} \sum_{k=1}^{K} \alpha_k \log\left(1 + \frac{|\boldsymbol{g}_k^H \boldsymbol{x}_k|^2}{\sum_{j \neq k} |\boldsymbol{g}_k^H \boldsymbol{x}_j|^2 + \sigma_R^2 \|\boldsymbol{g}_k^H \mathbf{F}\|^2 + \sigma_k^2}\right)$$
$$- P_\rho(\mathcal{X})$$
$$\text{s.t.} \ \mathrm{Tr}(\bar{\mathbf{V}}\bar{\mathbf{V}}^H) \leq P_S, \quad (12)$$
$$\|\bar{\mathbf{X}}\|_F^2 + \sigma_R^2 \|\bar{\mathbf{F}}\|_F^2 \leq P_R$$

where

$$P_\rho(\mathcal{X}) \triangleq \frac{1}{2\rho} \left( \|\mathbf{X} - \mathbf{FHV} + \rho \mathbf{Z}\|^2 + \|\sigma_R \mathbf{F} - \sigma_R \bar{\mathbf{F}} + \rho \mathbf{Z}_f\|^2 \right.$$
$$\left. + \|\mathbf{X} - \bar{\mathbf{X}} + \rho \mathbf{Z}_x\|^2 + \|\mathbf{V} - \bar{\mathbf{V}} + \rho \mathbf{Z}_v\|^2 \right),$$

and $\mathbf{Z}$, $\mathbf{Z}_f$, $\mathbf{Z}_x$ and $\mathbf{Z}_v$ are the dual variables associated with the equality constraints of problem (11).

Next, we show how to solve problem (12) using BSUM. The key to apply BSUM to (12) is to find a tractable locally tight lower bound for the objective of (12). To do so, we resort to the well-known WMMSE method [47]. First, by the theory of the WMMSE method, we have the following lemma.

*Lemma 3.1:* For each $k$, we have

$$\log\left(1 + \frac{|\boldsymbol{g}_k^H \boldsymbol{x}_k|^2}{\sum_{j \neq k} |\boldsymbol{g}_k^H \boldsymbol{x}_j|^2 + \sigma_R^2 \|\boldsymbol{g}_k^H \mathbf{F}\|^2 + \sigma_k^2}\right) \quad (13)$$
$$= \max_{u_k, w_k} \log(w_k) - w_k e_k(u_k, \mathbf{X}, \mathbf{F}) + 1$$

where $e_k(u_k, \mathbf{X}, \mathbf{F}) \triangleq |1 - u_k^* \boldsymbol{g}_k^H \boldsymbol{x}_k|^2 + \sum_{j \neq k} \alpha_k |u_k^* \boldsymbol{g}_k^H \boldsymbol{x}_j|^2 + \sigma_R^2 \|u_k^* \boldsymbol{g}_k^H \mathbf{F}\|^2 + \sigma_k^2 |u_k|^2$.

This lemma can be easily proven by checking the first-order optimality condition of the problem on the right-hand-side (rhs) of (13), leading to the optimal $u_k$ and $w_k$ (given $\mathbf{X}$ and $\mathbf{F}$) as follows

$$u_k(\mathbf{X}, \mathbf{F}) = \frac{\boldsymbol{g}_k^H \boldsymbol{x}_k}{\sum_{k=1}^{K} |\boldsymbol{g}_k^H \boldsymbol{x}_j|^2 + \sigma_R^2 \|\boldsymbol{g}_k^H \mathbf{F}\|^2 + \sigma_k^2}, \quad (14)$$

$$w_k(\mathbf{X}, \mathbf{F}) = \frac{1}{e_k(u_k(\mathbf{X}, \mathbf{F}), \mathbf{X}, \mathbf{F})}$$
$$= \frac{1}{1 - u_k^*(\mathbf{X}, \mathbf{F}) \boldsymbol{g}_k^H \boldsymbol{x}_k} \quad (15)$$
$$= 1 + \frac{|\boldsymbol{g}_k^H \boldsymbol{x}_k|^2}{\sum_{j \neq k} |\boldsymbol{g}_k^H \boldsymbol{x}_j|^2 + \sigma_R^2 \|\boldsymbol{g}_k^H \mathbf{F}\|^2 + \sigma_k^2},$$

where we have denoted the optimal $u_k$ and $w_k$ as $u_k(\mathbf{X}, \mathbf{F})$ and $w_k(\mathbf{X}, \mathbf{F})$ for a clear illustration of their dependence on $\mathbf{X}$ and $\mathbf{F}$. As a direct result of Lemma 3.1, we have

$$\log\left(1 + \frac{|\boldsymbol{g}_k^H \boldsymbol{x}_k|^2}{\sum_{j \neq k} |\boldsymbol{g}_k^H \boldsymbol{x}_j|^2 + \sigma_R^2 \|\boldsymbol{g}_k^H \mathbf{F}\|^2 + \sigma_k^2}\right)$$
$$\geq \log(w_k(\tilde{\mathbf{X}}, \tilde{\mathbf{F}})) - w_k(\tilde{\mathbf{X}}, \tilde{\mathbf{F}}) e_k(u_k(\tilde{\mathbf{X}}, \tilde{\mathbf{F}}), \mathbf{X}, \mathbf{F}) + 1, \quad (16)$$
$$\forall \tilde{\mathbf{X}}, \tilde{\mathbf{F}}, \mathbf{X}, \mathbf{F}.$$

Moreover, it can be easily verified that the rhs of (16) is a locally tight lower bound of the rate function shown on the lhs of (16). With such a tractable locally tight lower bound, we can easily apply BSUM to (12) with the block variables separated as 1) $\mathbf{F}$, 2) $\mathbf{X}$, 3) $\{\bar{\mathbf{V}}, \bar{\mathbf{X}}, \bar{\mathbf{F}}\}$, and 4) $\mathbf{V}$. Specifically, by applying the lower bound shown in (16), we propose to solve the following problem

$$\min_{\mathcal{X}} \sum_{k=1}^{K} w_k \alpha_k e_k(u_k, \mathbf{X}, \mathbf{F}) + P_\rho(\mathcal{X})$$
$$\text{s.t. } \text{Tr}\left(\bar{\mathbf{V}}\bar{\mathbf{V}}^H\right) \leq P_S, \quad (17)$$
$$\left\|\bar{\mathbf{X}}\right\|_F^2 + \sigma^2 \left\|\bar{\mathbf{F}}\right\|_F^2 \leq P_R$$

where $w_k$ and $u_k$ are given. Further, by simple manipulations, we can rewrite the above problem compactly as

$$\min_{\mathcal{X}} \text{Tr}\left(\mathbf{X}^H \mathbf{G}_w \mathbf{X}\right) - 2\Re e\left\{\text{Tr}\left(\mathbf{X}^H \mathbf{G}\mathbf{D}_w\right)\right\}$$
$$+ \sigma_R^2 \text{Tr}\left(\mathbf{F}^H \mathbf{G}_w \mathbf{F}\right) + P_\rho(\mathcal{X}) \quad (18)$$
$$\text{s.t. } \text{Tr}\left(\bar{\mathbf{V}}\bar{\mathbf{V}}^H\right) \leq P_S,$$
$$\left\|\bar{\mathbf{X}}\right\|_F^2 + \sigma_R^2 \left\|\bar{\mathbf{F}}\right\|_F^2 \leq P_R$$

where

$$\mathbf{G}_w \triangleq \sum_{k=1}^{K} w_k \alpha_k |u_k|^2 \boldsymbol{g}_k \boldsymbol{g}_k^H \text{ and } \mathbf{D}_w \triangleq \text{diag}\left\{(w_k \alpha_k u_k)_k\right\}.$$
$$(19)$$

Now we are ready to show the BSUM iteration for problem (18), which consists of the following four steps.

*1) Step 1: solving* (18) *for* $\mathbf{F}$ *given* $\{\mathbf{V}, \bar{\mathbf{F}}\}$: The $\mathbf{F}$-subproblem is an unconstrained quadratic optimization problem. By the first order optimality condition, we obtain

$$\sigma_R^2(2\rho \mathbf{G}_w + \mathbf{I})\mathbf{F} + \mathbf{F}\mathbf{H}\mathbf{V}\mathbf{V}^H\mathbf{H}^H$$
$$= \sigma_R(\sigma_R \bar{\mathbf{F}} - \rho \mathbf{Z}_f) + (\mathbf{X} + \rho \mathbf{Z})\mathbf{V}^H\mathbf{H}^H \quad (20)$$

which is the so-called Sylvester equation and admits efficient unique solution [48].

*2) Step 2: solving* (18) *for* $\{\bar{\mathbf{X}}, \bar{\mathbf{F}}, \bar{\mathbf{V}}\}$ *given* $\{\mathbf{X}, \mathbf{F}\}$: The subproblem with respect to $\{\bar{\mathbf{X}}, \bar{\mathbf{F}}, \bar{\mathbf{V}}\}$ can be further divided into two *independent* problems: one is with respect to $\bar{\mathbf{V}}$ while the other is with respect to $\{\bar{\mathbf{F}}, \bar{\mathbf{V}}\}$. Both problems are equivalent to projection of a point onto a ball centered at the origin, which can be solved in closed-form. Specifically, we have

$$\bar{\mathbf{V}} = \mathcal{P}_{P_S}\{\mathbf{V} + \rho \mathbf{Z}_v\}, \quad (21)$$
$$[\bar{\mathbf{X}} \quad \sigma_R \bar{\mathbf{F}}] = \mathcal{P}_{P_R}\{[\mathbf{X} + \rho \mathbf{Z}_x \quad \sigma_R \mathbf{F} + \rho \mathbf{Z}_f]\}. \quad (22)$$

where $\mathcal{P}_{\mathcal{X}}(\boldsymbol{x})$ denotes the projection of $\boldsymbol{x}$ onto the convex set $\mathcal{X}$. From (22), we can obtain the optimal $\bar{\mathbf{F}}$. It is worth mentioning that, the $\sigma_R$ in the constraint $\sigma_R \mathbf{F} = \sigma_R \bar{\mathbf{F}}$ is introduced to make the subproblem with respect to $\{\bar{\mathbf{F}}, \bar{\mathbf{X}}\}$ have a closed-form solution; otherwise, we need to solve a quadratic equation to get the optimal Lagrange multiplier associated with the relay power constraint.

TABLE III
ALGORITHM 2: BSUM ALGORITHM FOR PROBLEM (12)

| |
| --- |
| 0.  initialize $\{\mathbf{F}, \mathbf{V}\}$ such that the power constraints |
| 1.  set $\mathbf{X} = \mathbf{FHV}$, $\bar{\mathbf{X}} = \mathbf{X}$, $\bar{\mathbf{F}} = \mathbf{F}$, $\bar{\mathbf{V}} = \mathbf{V}$ |
| 2.  **repeat** |
| 3.    compute $\boldsymbol{u}$ and $\boldsymbol{w}$ via (14) and (15) |
| 4.    compute $\mathbf{G}_w$ and $\mathbf{D}_w$ via (19) |
| 5.    update $\mathbf{F}$ by solving Eq. (20) |
| 6.    update $\bar{\mathbf{V}}$ via Eq. (21) |
| 7.    update $\bar{\mathbf{X}}$ and $\bar{\mathbf{F}}$ via Eq. (22) |
| 8.    update $\mathbf{X}$ via (23) |
| 9.    update $\mathbf{V}$ via (24) |
| 10. **until** some termination criterion is met |

*3) Step 3: solving* (18) *for* $\mathbf{X}$ *given* $\{\bar{\mathbf{X}}, \mathbf{V}, \bar{\mathbf{F}}\}$: The $\mathbf{X}$-subproblem is also an unconstrained quadratic optimization problem. Again, by the first order optimality condition, we obtain a unique closed-form solution as follows

$$\mathbf{X} = \frac{1}{2}(\rho \mathbf{G}_w + \mathbf{I})^{-1}(2\rho \mathbf{G}\mathbf{D}_w + (\mathbf{FHV} - \rho \mathbf{Z}) + (\bar{\mathbf{X}} - \rho \mathbf{Z}_x)). \quad (23)$$

*4) Step 4: solving* (18) *for* $\mathbf{V}$ *given* $\{\bar{\mathbf{V}}, \mathbf{X}, \bar{\mathbf{F}}\}$: The $\mathbf{V}$-subproblem is also an unconstrained quadratic optimization problem. Similarly, we obtain a unique closed-form solution as follows by applying the first-order optimality condition

$$\mathbf{V} = (\mathbf{I} + \mathbf{H}^H \mathbf{F}^H \mathbf{FH})^{-1}(\bar{\mathbf{V}} - \rho \mathbf{Z}_v + \mathbf{H}^H \mathbf{F}^H(\mathbf{X} + \rho \mathbf{Z})). \quad (24)$$

In sum, every step of the BSUM iteration has a unique closed-form solution. Combining the steps for computing the lower bound, i.e., (14) and (15), we summarize the BSUM algorithm for (12) in Table III.

*C. Numerical results*

This section presents some numerical results to evaluate the performance of the proposed PDD method by comparing the alternating optimization (AO) method. In the AO method applied to problem $(P2)$, we alternatingly update the source precoder $\mathbf{V}$ and $\mathbf{F}$ while fixing the other. Specifically, in each iteration of the AO method, after updating $\boldsymbol{u}$ and $\boldsymbol{w}$ via (14) and (15), we alternatingly optimize $\mathbf{F}$ and $\mathbf{V}$ by solving the following problem for the optimized variable

$$\min \text{Tr}\left(\mathbf{V}^H \mathbf{H}^H \mathbf{F}^H \mathbf{G}_w \mathbf{FHV}\right)$$
$$- 2\Re e\left\{\text{Tr}\left(\mathbf{V}^H \mathbf{H}^H \mathbf{F}^H \mathbf{GD}_w\right)\right\} + \sigma_R^2 \text{Tr}\left(\mathbf{F}^H \mathbf{G}_w \mathbf{F}\right) \quad (25)$$
$$\text{s.t. } \text{Tr}\left(\mathbf{VV}^H\right) \leq P_S,$$
$$\left\|\mathbf{FHV}\right\|_F^2 + \sigma_R^2 \left\|\mathbf{F}\right\|_F^2 \leq P_R$$

leading to $\mathbf{F}$-subproblem and $\mathbf{V}$-subproblem. The $\mathbf{F}$-subproblem can be solved using Bisection method while the $\mathbf{V}$-subproblem can be solved by interior-point method, e.g., using the off-the-shelf package SeDuMi. Due to the coupling between the source precoder and the relay precoder in the relay power constraint, the AO method is not necessarily converge to KKT solutions of problem $(P2)$. Specifically, for this problem, the AO method can easily get trapped in some inefficient feasible point, as shown below.

In the simulations, we set $\alpha_k = 1, \forall k$, and the noise power is set to unit for all receivers (i.e., $\sigma_k^2 = \sigma_R^2 = 1, \forall k$). It is further assumed that the relay and the source have the same power budget $P$ for simplicity, i.e., $P_S = P_R = P$, and define $SNR \triangleq 10\log_{10}(P)$. Furthermore, each channel coefficient in both $\mathbf{H}$ and $\mathbf{G}$ is generated from the zero mean complex Gaussian distribution with unit variance. Moreover, for convenience, we denote by $(N_s, N_r, K)$ a relay BC network with $N_s$ source antenna, $N_r$ relay antennas, and $K$ users. For the PDD method, we set the initial penalty parameter $\rho_0 = \frac{500K}{2KN_r + N_r^2 + KN_s}$.

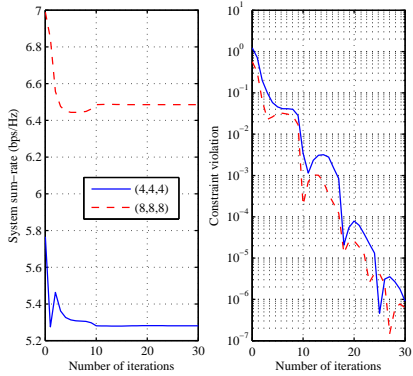

Fig. 3. The convergence performance of the PDD method for different relay BC networks.

The average convergence behavior of the PDD method over ten randomly generated examples is illustrated in Fig. 3, compared with the AO method in Fig. 4. It is observed from Fig. 3 that the PDD method exhibits excellent convergence performance in terms of both the objective value and the feasibility gap. In general, the PDD can converge in 20 iterations for both the network $(4, 4, 4)$ and the network $(8, 8, 8)$. Moreover, it is seen from Fig. 4 that the AO method is not only slow but also gets trapped in inefficient solutions whose objective values are much smaller than that achieved by the PDD method (i.e., right Y-axis versus left Y-axis).

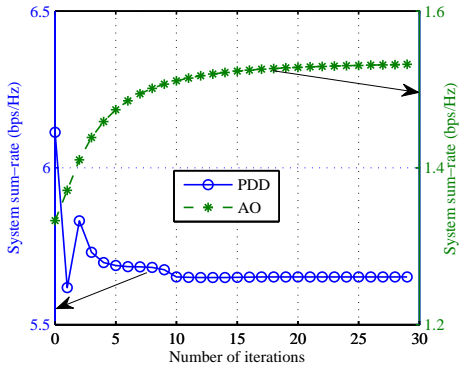

Fig. 4. The convergence performance of the PDD method and the AO method for network $(4, 4, 4)$.

Fig. 5 shows the average sum-rate performance of the PDD method as compared to the AO method for three different networks. Each result of the plot is averaged over 100 channels. One can see that, the PDD method always significantly outperforms the AO mehtod. his is mainly due to the fac that the AO method gets often trapped in some inefficient solutions due to the coupling between the source precoder and the relay precoder. In particular, as the variable dimension grows with the network size, the nonlinear coupling between the variables becomes more heavy and the AO method exhibits worse performance. In addition, we find that the PDD method is always more efficient than the AO method in terms of the cpu time required for convergence, as shown in the caption of Fig. 5.

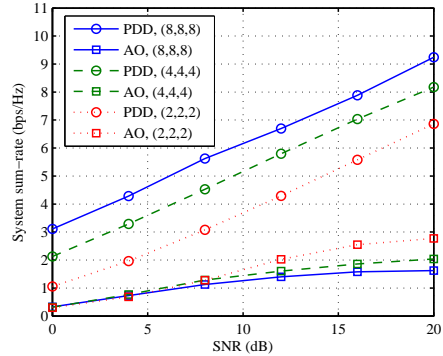

Fig. 5. The sum-rate performance of the PDD method and the AO method. The average ratio between the cpu time required by the AO method and by the PDD method is respectively 26.83, 39.06, and 12.82 for three networks.

## IV. VOLMIN-BASED MATRIX FACTORIZATION

As a popular tool in signal processing and machine learning, matrix factorization (MF) has attracted considerable interest in recent years. In addition to the most popular nonnegative matrix factorization (NMF) [37], various matrix factorization models have been proposed in the literature. Among them, volume-minimization (VolMin)-based matrix factorization is an important class of matrix factorizations where the columns of one factor matrix are constrained to lie in the unit simplex [36]. Compared to NMF, VolMin-based matrix factorization is computationally more challenging. This section considers application of PDD to VolMin-based matrix factorization and provides an alterative VolMin algorithm which can work in the original data space.

### A. Problem formulation

Consider the following data measurement model:

$$\boldsymbol{a}[\ell] = \mathbf{X}\boldsymbol{s}[\ell], \ell = 1, 2, \ldots, L, \quad (26)$$

where $\boldsymbol{a}[\ell] \in \mathbb{R}^N$ is a measured data vector indexed by $\ell$, $\mathbf{X} \in \mathbb{R}^{N \times K}$ denotes a basis which is assumed to have full column-rank, $\boldsymbol{s}[\ell] \in \mathbb{R}^K$ is the weight vector lying in a probability simplex, i.e.,

$$\boldsymbol{s}[\ell] \geq 0, \mathbf{1}^T \boldsymbol{s}[\ell] = 1, \forall \ell. \quad (27)$$

Define $\mathbf{S} \triangleq [\boldsymbol{s}[1]\ \boldsymbol{s}[2]\ \ldots\ \boldsymbol{s}[L]]$ and $\mathbf{A} \triangleq [\boldsymbol{a}[1]\ \boldsymbol{a}[2]\ \ldots\ \boldsymbol{a}[L]]$. Then the signal model (26) can be compactly written as

$$\mathbf{A} = \mathbf{X}\mathbf{S}. \quad (28)$$

An important motivating example of this model is hyperspectral remote sensing [6], where $\boldsymbol{a}[\ell]$ represents a remotely sensed pixel using sensors of high spectral resolution, the columns of $\mathbf{X}$ denote $K$ different spectral signatures of materials that comprise the pixels, $\boldsymbol{s}_k[\ell]$ denotes the portion of material $k$ contained in pixel $\boldsymbol{x}[\ell]$. Recovering $\mathbf{X}$ is helpful in recognition of the underlying materials in a hyperspectral image. Other applications of this model can be found in document clustering, multi-sensor array processing and blind separation of power spectra for dynamic spectrum access [7], [38], [49].

Given the data measurements $\mathbf{A}$, there possibly exist many combinations of factors $\mathbf{X}$ and $\mathbf{S}$ such $\mathbf{A} = \mathbf{XS}$. The notable works [7] and [50] showed that, under some realistic conditions, unique loading factors (up to column permutations) can be obtained by finding a minimum-volume enclosing simplex of the data vectors. Formally, the VolMin problem can be formulated as[3] [7]

$$\min_{\mathbf{X},\mathbf{S}} \ \log\det(\mathbf{X}^T\mathbf{X})$$
$$\text{s.t. } \mathbf{A} = \mathbf{XS}, \quad (29)$$
$$\mathbf{S}^T\mathbf{1} = \mathbf{1}, \mathbf{S} \geq 0.$$

Problem (29) is challenging due to the nonconvex objective function and the presence of the coupling constraint $\mathbf{A}=\mathbf{XS}$. Moreover, it is readily seen that the objective function is not well-defined for rank-deficient $\mathbf{X}$, which may produce numerical instability for iterative algorithms that cannot guarantee full-rank of $\mathbf{X}$ during iterations. To make it well-defined, we modify the objective function as $f_\epsilon(\mathbf{X}^T\mathbf{X})$, defined by

$$f_\epsilon(\mathbf{Y}) \triangleq \sum_{i=1}^n \log(g_\epsilon(\sigma_i(\mathbf{Y})))$$

where $g_\epsilon(\cdot)$ is given by

$$g_\epsilon(x) \triangleq \begin{cases} x & \text{if } |x| \geq \epsilon \\ \frac{1}{2\epsilon}x^2 + \frac{\epsilon}{2} & \text{otherwise} \end{cases} \quad (30)$$

where $\epsilon$ is a small scalar, e.g., $1e-2$, to control the approximation accuracy. With the modified objective function, we obtain an approximation of problem (29)

$$\boxed{\begin{aligned}\min_{\mathbf{X},\mathbf{S}} \ & f_\epsilon(\mathbf{X}^T\mathbf{X}) \\ \text{s.t. } & \mathbf{A} = \mathbf{XS}, \\ & \mathbf{S}^T\mathbf{1} = \mathbf{1}, \mathbf{S} \geq 0.\end{aligned}} \quad (P3)$$

It is noted from (30) that $f_\epsilon(\mathbf{X}^T\mathbf{X}) = \log\det(\mathbf{X}^T\mathbf{X})$ when the smallest singular value of $\mathbf{X}$ is equal or larger than $\sqrt{\epsilon}$. As a result, problem (P3) is equivalent to problem (29) when all the singular values of the optimal solution $\mathbf{X}$ to problem (29) are no smaller than $\sqrt{\epsilon}$. Note that, since the singular value of the optimal $\mathbf{X}$ can be easily ensured to be larger than $\sqrt{\epsilon}$

[3]It is worth mentioning that, we can also apply the PDD to an equivalent problem of (29) in the reduced-dimension domain (see (45) below). However, we here focus on the original data domain because it allows incorporating constraints on $\mathbf{X}$ (e.g., nonnegativity constraints or other box constraints) into (29) easily, and aims to shed lights on algorithm design for VolMin-related problems in the original data domain.

by appropriately scaling up $\mathbf{A}$, the approximation (P3) could incur no loss of optimality.

### B. PDD-based algorithm

In this subsection, we develop PDD-based algorithm to address problem $(P3)$. First we reformulate $(P3)$ as follows

$$\min_{\mathbf{X},\mathbf{S},\mathbf{Y}} \ f_\epsilon(\mathbf{X}^T\mathbf{X})$$
$$\text{s.t. } \mathbf{A} = \mathbf{YS} \quad (31)$$
$$\mathbf{X} - \mathbf{Y} = 0,$$
$$\mathbf{S}^T\mathbf{1} = \mathbf{1}, \mathbf{S} \geq 0.$$

By building the first two equality constraints into the objective, we obtain the augmented Lagrangian problem of the above reformulation as follows

$$\min_{\mathbf{X},\mathbf{S},\mathbf{Y}} \ f_\epsilon(\mathbf{X}^T\mathbf{X}) + \frac{1}{2\rho}\|\mathbf{A} + \rho\mathbf{P} - \mathbf{YS}\|^2$$
$$\quad\quad\quad + \frac{1}{2\rho}\|\mathbf{X} + \rho\mathbf{Q} - \mathbf{Y}\|^2 \quad (32)$$
$$\text{s.t. } \mathbf{S}^T\mathbf{1} = \mathbf{1}, \mathbf{S} \geq 0.$$

where $\mathbf{P}$ and $\mathbf{Q}$ are the Lagrange multipliers associated with the first two equality constraints of problem (31), respectively.

Next, we present the BSUM algorithm for problem (32), which consists of the following four steps.

*1) Step 1: Update $\mathbf{Y}$ given $\mathbf{X}$ and $\mathbf{S}$:* : Fixing $\mathbf{X}$ and $\mathbf{S}$ in (32), we obtain the subproblem with respect to $\mathbf{Y}$ as follows

$$\min_{\mathbf{Y}} \ \|\mathbf{A} + \rho\mathbf{P} - \mathbf{YS}\|^2 + \|\mathbf{X} + \rho\mathbf{Q} - \mathbf{Y}\|^2 \quad (33)$$

It is a quadratic optimization problem which admits a closed-form solution as follows

$$\mathbf{Y} = \left((\mathbf{A} + \rho\mathbf{P})\mathbf{S}^T + (\mathbf{X} + \rho\mathbf{Q})\right)(\mathbf{I} + \mathbf{SS}^T)^{-1}. \quad (34)$$

*2) Step 2: Update $\mathbf{S}$ given $\mathbf{X}$ and $\mathbf{Y}$:* : By fixing $\mathbf{X}$ and $\mathbf{Y}$ in (32), we obtain the subproblem with respect to $\mathbf{S}$ as follows

$$\min_{\mathbf{S}} \ \|\mathbf{YS} - (\mathbf{A} + \rho\mathbf{P})\|^2$$
$$\text{s.t. } \mathbf{1}^T\mathbf{S} = \mathbf{1}, \mathbf{S} \geq 0. \quad (35)$$

The above problem is a convex problem which can be globally solved by using some iterative algorithms. To obtain an efficient update for $\mathbf{S}$, we consider updating $\mathbf{S}$ by minimizing a locally tight upper bound of the objective function of (35), i.e., solving

$$\min_{\mathbf{S}} \ \|\mathbf{YS} - (\mathbf{A} + \rho\mathbf{P})\|^2 + \|\mathbf{S} - \tilde{\mathbf{S}}\|_{\mathbf{W}}^2$$
$$\text{s.t. } \mathbf{1}^T\mathbf{S} = \mathbf{1}, \mathbf{S} \geq 0. \quad (36)$$

where $\tilde{\mathbf{S}}$ is the value of $\mathbf{S}$ obtained in the last iteration and $\mathbf{W}$ is a positive definite matrix such that $\mathbf{Y} + \mathbf{W} = \beta\mathbf{I}$ with $\beta > (\sigma_1(\mathbf{Y}))^2$. With simple manipulations, problem (36) can be equivalently written as

$$\min_{\mathbf{S}} \ \|\mathbf{S} - \bar{\mathbf{S}}\|^2,$$
$$\text{s.t. } \mathbf{1}^T\mathbf{S} = \mathbf{1}, \mathbf{S} \geq 0. \quad (37)$$

where $\bar{\mathbf{S}} = \frac{1}{\beta}\left(\mathbf{Y}^T(\mathbf{A} + \rho\mathbf{P}) + (\beta\mathbf{I} - \mathbf{Y}^T\mathbf{Y})\tilde{\mathbf{S}}\right)$. Problem (37) can be decomposed into $L$ independent subproblems which are known as the problem of *projection onto the probability simplex* and admit very efficient semi-closed-form solutions (see [51]).

*3) Update $\mathbf{X}$ given $\mathbf{Y}$ and $\mathbf{S}$:* By fixing $\mathbf{Y}$ and $\mathbf{S}$ in (32), we obtain the subproblem with respect to $\mathbf{X}$ as follows

$$\min_{\mathbf{X}} \; f_\epsilon(\mathbf{X}^T\mathbf{X}) + \frac{1}{2\rho}\|\mathbf{X} - \bar{\mathbf{X}}\|^2 \tag{38}$$

where $\bar{\mathbf{X}} \triangleq \mathbf{Y} - \rho\mathbf{Q}$. By the definition of $f_\epsilon(\cdot)$, it is known that $f_\epsilon(\mathbf{X}^T\mathbf{X})$ is only related to the singular values of $\mathbf{X}$. Furthermore, by *Von Neumann's trace inequality* $|\text{Tr}(\mathbf{X}^T\bar{\mathbf{X}})| \le \sum_{i=1}^n \sigma_i(\mathbf{X})\sigma_i(\bar{\mathbf{X}})$, it is easily known that the singular vectors of the optimal $\mathbf{X}$ should be aligned with those of $\bar{\mathbf{X}}$. Hence, letting $\bar{\mathbf{U}}\bar{\mathbf{\Sigma}}\bar{\mathbf{V}}^H$ be the thin SVD of $\bar{\mathbf{X}}$, the optimal $\mathbf{X}$ is structured as $\mathbf{X} = \bar{\mathbf{U}}\mathbf{\Sigma}\bar{\mathbf{V}}^H$. As a result, problem (38) reduces to

$$\min_{\mathbf{\Sigma}\ge 0} \; f_\epsilon(\mathbf{\Sigma}^2) + \frac{1}{2\rho}\|\mathbf{\Sigma} - \bar{\mathbf{\Sigma}}\|^2 \tag{39}$$

which can be decomposed into a set of independent subproblems with the $i$-th subproblem in the form of

$$\min_{\sigma_i \ge 0} \; \log(g_\epsilon(\sigma_i^2)) + \frac{1}{2\rho}(\sigma_i - \bar{\sigma}_i)^2 \tag{40}$$

where $\sigma_i$ is the $i$-th singular value of $\mathbf{X}$. The above problem is difficult to solve and thus we devote our efforts to minimizing a locally tight upper bound of the objective function. By the concavity of the $\log(\cdot)$ function, we have $\log(x) \le \log(\tilde{x}) + \frac{1}{\tilde{x}}(x - \tilde{x}), \forall x, \tilde{x} > 0$. Using such an upper bound for $\log(g_\epsilon(\sigma_i^2))$, we update $\sigma_i$ by solving

$$\min_{\sigma_i \ge 0} \; \frac{1}{\tilde{g}_\epsilon} g_\epsilon(\sigma_i^2) + \frac{1}{2\rho}(\sigma_i - \bar{\sigma}_i)^2 \tag{41}$$

where $\tilde{g}_\epsilon = g_\epsilon\left((\tilde{\sigma}_i)^2\right)$ and $\tilde{\sigma}_i = \sigma_i(\tilde{\mathbf{X}})$ with $\tilde{\mathbf{X}}$ being the value of $\mathbf{X}$ obtained in the last iteration.

It can be shown that the objective function of problem (41) is convex with respect to $\sigma_i \ge 0$. Hence, problem (41) is a convex problem and can be globally solved. Specifically, we solve it by considering two cases. The first case is when $\sigma_i^2 \ge \epsilon$. In this case, problem (41) reduces to

$$\min_{\sigma_i \ge \sqrt{\epsilon}} \; \frac{1}{\tilde{g}_\epsilon}\sigma_i^2 + \frac{1}{2\rho}(\sigma_i - \bar{\sigma}_i)^2 \tag{42}$$

which admits a closed-form solution as follows

$$\sigma_i = \max\left(\frac{\tilde{g}_\epsilon}{2\rho + \tilde{g}_\epsilon}\bar{\sigma}_i, \sqrt{\epsilon}\right)$$

In the second case when $\sigma_i^2 \le \epsilon$, problem (41) reduces to

$$\min_{0 \le \sigma_i \le \sqrt{\epsilon}} \; \frac{1}{2\epsilon\tilde{g}_\epsilon}\sigma_i^4 + \frac{1}{2\rho_k}(\sigma_i - \bar{\sigma}_i)^2 \tag{43}$$

which admits a closed-form solution as $\sigma_i = [\sigma_i^\star]_0^{\sqrt{\epsilon}}$ where $\sigma_i^\star$ is the unique solution to the following cubic equation

$$\frac{2}{\epsilon\tilde{g}_\epsilon}\sigma_i^3 + \frac{1}{\rho_k}\sigma_i - \frac{1}{\rho_k}\bar{\sigma}_i = 0.$$

By comparing the objective values of the above two cases, we can obtain the optimal $\sigma_i$ for problem (41). After obtaining $\mathbf{\Sigma}$, we finally obtain the optimal solution $\mathbf{X}$ to problem (38), i.e., $\mathbf{X} = \bar{\mathbf{U}}\mathbf{\Sigma}\bar{\mathbf{V}}^H$. We omit the detailed implementation of the BSUM algorithm for problem (32) due to space limitation.

*4) Numerical examples:* We here present numerical examples to illustrate the performance of the PDD-based VolMin algorithm by comparing with the state-of-art VolMin algorithm—SISAL [6]. SISAL works for the equivalent problem of problem ($P3$) in the reduced-dimension domain [6],

$$\begin{aligned}\max_{\mathbf{Q}\in\mathbb{R}^{K\times K}} & \quad \log|\det(\mathbf{Q})| \\ \text{s.t.} & \quad \mathbf{Q}^T\mathbf{1} = (\mathbf{A}_r\mathbf{A}_r^T)^{-1}\mathbf{A}_r\mathbf{1}, \mathbf{Q}\mathbf{A}_r \ge 0.\end{aligned} \tag{44}$$

where $\mathbf{A}_r \triangleq \mathbf{U}_r^T\mathbf{A}$ and $\mathbf{U}_r \in \mathbb{R}^{N\times K}$ consists of the left-singular vectors of $\mathbf{A}$ (which can be obtained by performing thin SVD on $\mathbf{A}$). In SISAL, the inequality constraints are penalized to the objective by using *hinge-loss* function, leading to a penalized problem [6]

$$\begin{aligned}\max_{\mathbf{Q}\in\mathbb{R}^{K\times K}} & \quad \log|\det(\mathbf{Q})| - \eta\sum_{i,j}\max(-[\mathbf{Q}\mathbf{A}_r]_{ij}, 0) \\ \text{s.t.} & \quad \mathbf{Q}^T\mathbf{1} = (\mathbf{A}_r\mathbf{A}_r^T)^{-1}\mathbf{1}.\end{aligned} \tag{45}$$

where $\eta$ is a penalty parameter. The SISAL algorithm aims to solve the penalized problem by using successive second-order approximation and variable splitting technique[4]. The algorithm is lightweight but its convergence is unclear. Moreover, the SISAL algorithm does not apply to the original data space.

In our simulations, we generate the elements of $\mathbf{X}$ from the uniform distribution between zero and one, and generate $s[\ell]$ on the unit simplex and with $\max_i s_i[\ell] \le \gamma$, where $\gamma = 0.8$ is given, which results in a so-called 'no-pure-pixel case' in the context of remote sensing and is known to be challenging to handle [6], [49]. Moreover, we use the mean-square-error (MSE) of $\mathbf{X}$ as a measure of estimation performance (instead of achieved volume which is less physically meaningful), defined by

$$MSE = \min_{\boldsymbol{\pi}\in\Pi} \frac{1}{K}\sum_{k=1}^K \left\|\frac{\boldsymbol{x}_k}{\|\boldsymbol{x}_k\|} - \frac{\hat{\boldsymbol{x}}_{\boldsymbol{\pi}_k}}{\|\hat{\boldsymbol{x}}_{\boldsymbol{\pi}_k}\|}\right\|^2 \tag{46}$$

where $\Pi$ is the set of all permutations of $\{1, 2, \ldots, K\}$, and $\hat{\boldsymbol{x}}_k$ is the estimate of $\boldsymbol{x}_k$. For the PDD method, the initial penalty parameter $\rho_0$ is set to $L/100$.

We first show in Fig. 6 the convergence performance of the PDD method. In the plot, the results are averaged over ten randomly generated examples with random initialization for two cases: $(N, K, L) = (10, 3, 200)$ and $(N, K, L) = (50, 3, 2000)$. It can be observed that the PDD method achieve approximate feasibility in tens of iterations. Particularly, it can quickly reach a good estimation accuracy; the MSE could be less than $-35$ dB in twenty outer iterations.

We then use two illustrative examples to show the effectiveness of the proposed PDD-based VolMin algorithm. In these two examples, we again set $(N, K, L) = (10, 3, 200)$ and $(N, K, L) = (50, 3, 2000)$. To visualize the results, we project the data points, the ground truth, and the estimates to a two-dimensional plane. In Figs. 7 and 8, we see that, the PDD method can provide an estimate as good as the SISAL method's if the latter is particularly initialized from the estimate of VCA method [52]. Moreover, we observe from simulations that the PDD method is less sensitive to random initialization than the SISAl method, as shown in the Figs. 7-8 and also Fig. 9 below.

---
[4]The code of SISAL can be found from http://www.lx.it.pt/~bioucas/code.htm.

To further demonstrate the performance of the PDD method under random initialization, we randomly generate 100 examples and evaluate the estimation performance of various methods. Motivated by the observations from Fig. 6, we simply set the maximum outer iterations of the PDD method as 30 in this set of simulations. Moreover, to test the performance of the PDD method in a noisy environment, zero-mean white Gaussian noise $v[\ell]$ is added to each generated data. We define the signal-to-noise ratio (SNR) as $SNR = 10\log_{10}\left(\frac{\mathbb{E}\{\|\mathbf{X}s[\ell]\|^2\}}{\mathbb{E}\{\|v[\ell]\|^2\}}\right)$, and set $SNR = 40$ dB for each example. Fig. 9 illustrates the estimation performance of various methods for 100 examples. We see that, the PDD is much more robust to random initializations than SISAL. In addition to random initialization, the performance of SISAL is also impacted by the choice of $\eta$. Moreover, the PDD method[5] with three random initializations can provide very high estimation performance that is comparable with the SISAL method when the latter is set with a finely tuned penalty parameter $\eta$ and particularly initialized from VCA.

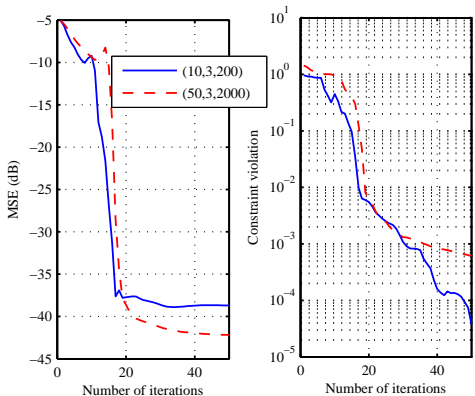

Fig. 6. The convergence performance of the PDD method.

## V. CONCLUSIONS

In this two-part paper, we proposed and analyzed a new optimization framework for optimizing nonconvex nonsmooth functions subject to nonconvex coupling constraints. Part I developed the general framework and investigated its convergence properties. In this Part II, we customized our PDD framework to three challenging problems in signal processing and machine learning. Our algorithms guarantee convergence to stationary solutions of the three problems[6] and were shown numerically to be able to yield better solutions than the state-of-the-art schemes in the literature. We remark that, our framework finds applications also in other areas, such as optimal power flow in smart grids, user scheduling in wireless communications, cross-layer design of networks, etc..

---

[5]To combat against the impact of initialization, we run PDD with three random initializations and pick the best one as the output.

[6]The verification of constraint qualification for the three problems is relegated to Appendix C.

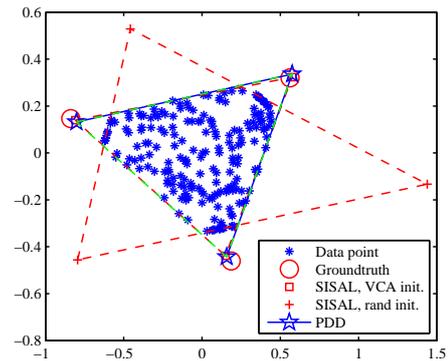

Fig. 7. A geometric illustration of estimation results by different methods with $N = 10$, $K = 3$, and $L = 200$.

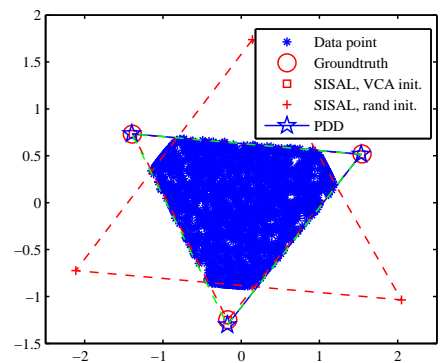

Fig. 8. A geometrical illustration of estimation results by different methods with $N = 50$, $K = 3$, and $L = 2000$.

## APPENDIX A
## SOLVING PROBLEM (6) FOR $t$

This appendix shows how the $t$-subproblem is globally solved. For a clear illustration, we write the $t$-subproblem

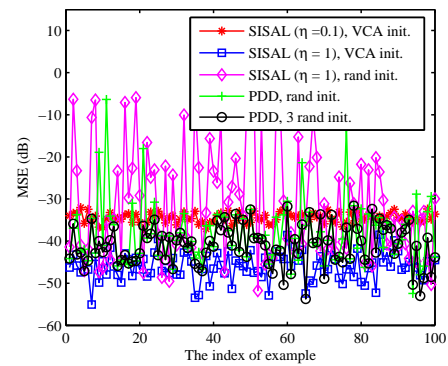

Fig. 9. The MSE performance of different methods with $N = 50$, $K = 3$, and $L = 1000$

explicitly as follows

$$\max_{\boldsymbol{t}\geq 0} \min_{1\leq k\leq K} t_k - \frac{1}{2\rho}\sum_{k=1}^{K}\left(\|\mathbf{A}_k^{\frac{1}{2}}\boldsymbol{w}\|-t_k\|\mathbf{B}_k^{\frac{1}{2}}\boldsymbol{w}\|+\rho\lambda_k\right)^2 \quad (47)$$

It can be compactly written as

$$\max_{\boldsymbol{t}\geq 0}\min_{k} t_k - \sum_{k=1}^{K} a_k(t_k - b_k)^2 \quad (48)$$

where $a_k \triangleq \frac{\|\mathbf{B}_k^{\frac{1}{2}}\boldsymbol{w}\|^2}{2\rho}$ and $b_k \triangleq \frac{\|\mathbf{A}_k^{\frac{1}{2}}\boldsymbol{w}\|+\rho\lambda_k}{\|\mathbf{B}_k^{\frac{1}{2}}\boldsymbol{w}\|}$. Further, by introducing an auxiliary variable $s$, the above problem can be equivalently written as

$$\max_{\boldsymbol{t}\geq 0, s} s - \sum_{k=1}^{K} a_k(t_k - b_k)^2 \quad (49)$$
$$\text{s.t. } t_k \geq s \geq 0, \forall k.$$

We solve the above problem by exploiting its problem structure. To do so, let $t_k^*$'s and $s^*$ denote the optimal solution to the above problem. It is seen from (49) that, if $t_k^* > s^* \geq 0$, then we must have $t_k^* = b_k$ to maximize the objective. Hence, by assuming without loss of generality $b_1 \geq b_2 \geq \ldots \geq b_K$, we infer that, there exists $\bar{k}$ such that $t_k^* = b_k$, $\forall k \leq \bar{k}$, and $t_k^* = s^*$, $\forall k > \bar{k}$, where $s^*$ is given by

$$s^* = \arg\max_{s\geq 0} s - \sum_{k>\bar{k}} a_k(s - b_k)^2 \quad (50)$$
$$= \max\left(\frac{1 + 2\sum_{k>\bar{k}} a_k b_k}{2\sum_{k>\bar{k}} a_k}, 0\right) \triangleq \tau(\bar{k}). \quad (51)$$

As a result, it is equivalent to find $\bar{k} > 0$ such that $b_{\bar{k}}^* > \tau(\bar{k})$ if such a $\bar{k}$ exists; otherwise we have $\bar{k} = 0$ (i.e., $t_k^* = s^* = \tau(0)$, $\forall k$). Given $\bar{k}$, we can derive the optimal solution as stated above.

## APPENDIX B
## THE EXPRESSION OF MATRIX $\mathbf{C}$

We here derive a homogeneous quadratic upper bound for $\vartheta(\boldsymbol{w})$ over the constraint $\|\boldsymbol{w}\| = 1$. Since $\vartheta(\boldsymbol{w})$ is in the form of

$$\vartheta(\boldsymbol{w}) = \sum_{k=1}^{K}\left(\|\mathbf{A}_k^{\frac{1}{2}}\boldsymbol{w}\|-t_k\|\mathbf{B}_k^{\frac{1}{2}}\boldsymbol{w}\|+\rho\lambda_k\right)^2,$$

we only need to bound each summand $\vartheta_k(\boldsymbol{w}) \triangleq \left(\|\mathbf{A}_k^{\frac{1}{2}}\boldsymbol{w}\|-t_k\|\mathbf{B}_k^{\frac{1}{2}}\boldsymbol{w}\|+\rho\lambda_k\right)^2$, $k = 1, 2, \ldots, K$.

First, $\vartheta_k(\boldsymbol{w})$ can be expressed as

$$\vartheta_k(\boldsymbol{w}) = \boldsymbol{w}^H\mathbf{A}_k\boldsymbol{w} + t_k^2\boldsymbol{w}^H\mathbf{B}_k\boldsymbol{w} + \rho^2\lambda_k^2 \\ - 2t_k\|\mathbf{A}_k^{\frac{1}{2}}\boldsymbol{w}\|\|\mathbf{B}_k^{\frac{1}{2}}\boldsymbol{w}\| + 2\rho g_k(\boldsymbol{w}) \quad (52)$$

where

$$g_k(\boldsymbol{w}) \triangleq \begin{cases} -\lambda_k t_k \|\boldsymbol{w}\|\left\|\mathbf{B}_k^{\frac{1}{2}}\boldsymbol{w}\right\| + \lambda_k\|\mathbf{A}_k^{\frac{1}{2}}\boldsymbol{w}\|, \text{if } \lambda_k \geq 0 \\ -\lambda_k t_k \left\|\mathbf{B}_k^{\frac{1}{2}}\boldsymbol{w}\right\| + \lambda_k\|\boldsymbol{w}\|\|\mathbf{A}_k^{\frac{1}{2}}\boldsymbol{w}\|, \text{otherwise} \end{cases} \quad (53)$$

Note that we have used the fact $\|\boldsymbol{w}\| = 1$ in the definition of $g_k(\boldsymbol{w})$, so that $g_k(\boldsymbol{w})$ has similar forms for both cases of $\lambda_k$, which can be easily upper bounded. In what follows, without loss of generality, we consider only the case when $\lambda_k \geq 0$.

To bound $\vartheta_k(\boldsymbol{w})$, let us define $\boldsymbol{w}_{eq} \triangleq (\Re e\{\boldsymbol{w}\}, \Im m\{\boldsymbol{w}\})$, $\tilde{\boldsymbol{w}}_{eq} \triangleq (\Re e\{\tilde{\boldsymbol{w}}\}, \Im m\{\tilde{\boldsymbol{w}}\})$, and

$$\mathbf{A}_{k,eq} \triangleq \begin{pmatrix} \Re e\{\mathbf{A}_k\} & -\Im m\{\mathbf{A}_k\} \\ \Im m\{\mathbf{A}_k\} & \Re e\{\mathbf{A}_k\} \end{pmatrix},$$
$$\mathbf{B}_{k,eq} \triangleq \begin{pmatrix} \Re e\{\mathbf{B}_k\} & -\Im m\{\mathbf{B}_k\} \\ \Im m\{\mathbf{B}_k\} & \Re e\{\mathbf{B}_k\} \end{pmatrix}.$$

Then, by applying part 1) of Lemma 2.1, we have

$$2t_k\left\|\mathbf{A}_k^{\frac{1}{2}}\boldsymbol{w}\right\|\left\|\mathbf{B}_k^{\frac{1}{2}}\boldsymbol{w}\right\| + 2\rho\lambda_k t_k\|\boldsymbol{w}\|\left\|\mathbf{B}_k^{\frac{1}{2}}\boldsymbol{w}\right\|$$
$$=2t_k\left\|\mathbf{A}_{k,eq}^{\frac{1}{2}}\boldsymbol{w}_{eq}\right\|\left\|\mathbf{B}_{k,eq}^{\frac{1}{2}}\boldsymbol{w}_{eq}\right\| + 2\rho\lambda_k t_k\|\boldsymbol{w}_{eq}\|\left\|\mathbf{B}_{k,eq}^{\frac{1}{2}}\boldsymbol{w}_{eq}\right\|$$
$$\geq \boldsymbol{w}_{eq}^T\boldsymbol{\Omega}_k(\tilde{\boldsymbol{w}})\boldsymbol{w}_{eq}$$

where

$$\boldsymbol{\Omega}_k(\tilde{\boldsymbol{w}}) \triangleq \frac{t_k}{\left\|\mathbf{A}_{k,eq}^{\frac{1}{2}}\tilde{\boldsymbol{w}}_{eq}\right\|\left\|\mathbf{B}_{k,eq}^{\frac{1}{2}}\tilde{\boldsymbol{w}}_{eq}\right\|}\left(\mathbf{A}_{k,eq}\tilde{\boldsymbol{w}}_{eq}\tilde{\boldsymbol{w}}_{eq}^T\mathbf{B}_{k,eq}\right.$$
$$\left.+\mathbf{B}_{k,eq}\tilde{\boldsymbol{w}}_{eq}\tilde{\boldsymbol{w}}_{eq}^T\mathbf{A}_{k,eq}\right)$$
$$+\frac{\rho\lambda_k t_k}{\|\tilde{\boldsymbol{w}}_{eq}\|\left\|\mathbf{B}_{k,eq}^{\frac{1}{2}}\tilde{\boldsymbol{w}}_{eq}\right\|}\left(\tilde{\boldsymbol{w}}_{eq}\tilde{\boldsymbol{w}}_{eq}^T\mathbf{B}_{k,eq} + \mathbf{B}_{k,eq}\tilde{\boldsymbol{w}}_{eq}\tilde{\boldsymbol{w}}_{eq}^T\right).$$
$$(54)$$

Furthermore, by applying part 2) of Lemma 2.1, we have

$$2\left\|\mathbf{A}_k^{\frac{1}{2}}\boldsymbol{w}\right\| = 2\left\|\mathbf{A}_{k,eq}^{\frac{1}{2}}\boldsymbol{w}_{eq}\right\|$$
$$\leq \frac{1}{\left\|\mathbf{A}_{k,eq}^{\frac{1}{2}}\tilde{\boldsymbol{w}}_{eq}\right\|}\left\|\mathbf{A}_{k,eq}^{\frac{1}{2}}\boldsymbol{w}_{eq}\right\|^2 + \left\|\mathbf{A}_{k,eq}^{\frac{1}{2}}\tilde{\boldsymbol{w}}_{eq}\right\| \quad (55)$$

As a result, we can obtain a locally tight quadratic upper bound for $\vartheta_k(\boldsymbol{w})$ given by

$$\vartheta_k(\boldsymbol{w}) \leq \boldsymbol{w}_{eq}^T\mathbf{C}_k\boldsymbol{w}_{eq} + const \quad (56)$$

where

$$\mathbf{C}_k \triangleq \left(1 + \frac{\rho\lambda_k}{\left\|\mathbf{A}_k^{\frac{1}{2}}\tilde{\boldsymbol{w}}\right\|}\right)\mathbf{A}_{k,eq} + t_k^2\mathbf{B}_{k,eq} - \boldsymbol{\Omega}_k(\tilde{\boldsymbol{w}}). \quad (57)$$

Finally, we have

$$\vartheta(\boldsymbol{w}) \leq u(\boldsymbol{w},\tilde{\boldsymbol{w}}) \triangleq \boldsymbol{w}_{eq}^T\mathbf{C}\boldsymbol{w}_{eq} + const \quad (58)$$

where $\mathbf{C} \triangleq \sum_{k=1}^{K}\mathbf{C}_k$.

## APPENDIX C
## CONSTRAINT QUALIFICATION OF PROBLEMS (5), (11), AND (31)

In this appendix, we verify the constraint qualification of problems (5), (11), and (31) by considering Mangasarian-Fromovitz constraint qualification (MFCQ) (which is equivalent to Robinson's condition for these problems).

First, let us introduce MFCQ for the constraints of the following problem

$$\min f(\boldsymbol{x})$$
$$\text{s.t. } \boldsymbol{h}(\boldsymbol{x}) = 0, \quad (59)$$
$$\boldsymbol{g}(\boldsymbol{x}) \leq 0,$$

where the functions $f : \mathbb{R}^n \to \mathbb{R}$, $g : \mathbb{R}^n \to \mathbb{R}^p$ and $h : \mathbb{R}^n \to \mathbb{R}^m$ are continuously differentiable. The feasible set is $\Omega = \{x \in \mathbb{R}^n \mid h(x) = 0, g(x) \leq 0\}$. Given $\bar{x} \in \Omega$, $\mathcal{A}(\bar{x})$ is the set of the inequality active constraint indices, that is

$$\mathcal{A}(\bar{x}) = \{i \in \{1, 2, \ldots, p\} \mid g_i(\bar{x}) = 0\}. \quad (60)$$

For problem (59), we say that *MFCQ holds at $\bar{x}$ when the equality constraint gradients are linearly independent and there exists a vector $d \in \mathbb{R}^n$* such that

$$\nabla h(\bar{x})d = 0, \quad (61)$$
$$\nabla g_j(\bar{x})^T d < 0, \forall j \in \mathcal{A}(\bar{x}). \quad (62)$$

Here, $\nabla h(\bar{x})$ denotes the Jacobian matrix of $h(x)$ and $\nabla g_j(\bar{x})$ is the gradient of $g_j(x)$. Thus, the equality constraint gradients are given by the columns of $\nabla h(\bar{x})^T$.

*Remark C.1:* By the first-order approximation, we have $h(\bar{x} + d) \approx h(\bar{x}) + \nabla h(\bar{x})d$. Hence, we can obtain the term $\nabla h(\bar{x})d$ by using first-order approximation without need of computing the Jacobian matrix (or the gradient). This observation will facilitate the MFCQ verification in the case when $h$ and $x$ are both matrices.

Next, let us check the MFCQ of three problems one by one.

### A. CQ verification for problem (5)

For problem (5), we have

*Lemma C.1:* MFCQ holds for problem (5) at any feasible point $(w, t)$.

*Proof:* It is readily seen that the inequality constraints $t \geq 0$ of problem (5) must be inactive. Hence, we only need to check the equality constraints which can be compactly written as $h(w, t) = 0$ with the following definition

$$h(w, t) \triangleq \begin{bmatrix} \|\mathbf{A}_1^{\frac{1}{2}} w\| - t_1 \|\mathbf{B}_1^{\frac{1}{2}} w\| \\ \|\mathbf{A}_2^{\frac{1}{2}} w\| - t_2 \|\mathbf{B}_2^{\frac{1}{2}} w\| \\ \vdots \\ \|w\|^2 - 1 \end{bmatrix}. \quad (63)$$

The equality constraint gradients are computed in (64) (see the top of the next page). By noting $w \neq 0$, it is readily known that the columns of $\nabla h(w,t)^T$ are linearly independent. Furthermore, by simply setting $d = 0$, MFCQ holds for problem (5) at any its feasible point. ∎

### B. CQ verification for problem (11)

For problem (11), we have

*Lemma C.2:* MFCQ holds for problem (11) at any *nonzero* feasible point $(\mathbf{V}, \mathbf{F}, \mathbf{X})$ with $\mathbf{V} \neq \mathbf{0}$, and $\mathbf{F} \neq \mathbf{0}$ or $\mathbf{X} \neq \mathbf{0}$.

*Proof:* The constraints of problem (11) are written as follows

$$g_1 \triangleq \mathrm{Tr}\left(\bar{\mathbf{V}}\bar{\mathbf{V}}^H\right) - P_S \leq 0, \quad (65a)$$
$$g_2 \triangleq \|\bar{\mathbf{X}}\|_F^2 + \sigma_R^2 \|\bar{\mathbf{F}}\|_F^2 - P_R \leq 0, \quad (65b)$$
$$\boldsymbol{\Theta}_1 \triangleq \mathbf{X} - \mathbf{FHV} = \mathbf{0}, \quad (65c)$$
$$\boldsymbol{\Theta}_2 \triangleq \mathbf{F} - \bar{\mathbf{F}} = \mathbf{0}, \quad (65d)$$
$$\boldsymbol{\Theta}_3 \triangleq \mathbf{X} - \bar{\mathbf{X}} = \mathbf{0}, \quad (65e)$$
$$\boldsymbol{\Theta}_4 \triangleq \mathbf{V} - \bar{\mathbf{V}} = \mathbf{0}. \quad (65f)$$

As we can see, $\boldsymbol{\Theta}_1$, $\boldsymbol{\Theta}_2$ and $\boldsymbol{\Theta}_3$ do not contain variable $\bar{\mathbf{V}}$ but $\boldsymbol{\Theta}_4$ does. Thus, the gradients of the components of $\boldsymbol{\Theta}_1$, $\boldsymbol{\Theta}_2$, $\boldsymbol{\Theta}_3$, and $\boldsymbol{\Theta}_4$ are linearly dependent if and only if those of the components of $\boldsymbol{\Theta}_1$, $\boldsymbol{\Theta}_2$, and $\boldsymbol{\Theta}_3$ are linearly dependent. Similarly, since $\boldsymbol{\Theta}_1$ and $\boldsymbol{\Theta}_2$ do not contain $\bar{\mathbf{X}}$, the gradients of the components of $\boldsymbol{\Theta}_1$, $\boldsymbol{\Theta}_2$, and $\boldsymbol{\Theta}_3$ are linearly dependent if and only if those of the components of $\boldsymbol{\Theta}_1$ and $\boldsymbol{\Theta}_2$ are linearly dependent. However, the gradients of the components of $\boldsymbol{\Theta}_1$ and $\boldsymbol{\Theta}_2$ are linearly independent because $\boldsymbol{\Theta}_1$ does not contain $\bar{\mathbf{F}}$ but $\boldsymbol{\Theta}_2$ does. Therefore, the equality constraint gradients of problem (11) are linearly independent.

Given the above gradient independence result, we are left to show that, there exists $\{\mathbf{D_X}, \mathbf{D_F}, \mathbf{D_V}, \mathbf{D_{\bar{X}}}, \mathbf{D_{\bar{F}}}, \mathbf{D_{\bar{V}}}\}$ such

$$\Re e\left\{\mathrm{Tr}\left(\bar{\mathbf{V}} \mathbf{D}_{\bar{\mathbf{V}}}^H\right)\right\} < 0, \quad (66a)$$
$$\Re e\left\{\mathrm{Tr}\left(\bar{\mathbf{X}} \mathbf{D}_{\bar{\mathbf{X}}}^H\right) + \sigma_R^2 \Re e\left\{\mathrm{Tr}\left(\bar{\mathbf{F}} \mathbf{D}_{\bar{\mathbf{F}}}^H\right)\right\}\right\} < 0, \quad (66b)$$
$$\mathbf{D_X} - \mathbf{D_F HV} - \mathbf{FH D_V} = \mathbf{0}, \quad (66c)$$
$$\mathbf{D_F} - \mathbf{D_{\bar{F}}} = \mathbf{0}, \quad (66d)$$
$$\mathbf{D_X} - \mathbf{D_{\bar{X}}} = \mathbf{0}, \quad (66e)$$
$$\mathbf{D_V} - \mathbf{D_{\bar{V}}} = \mathbf{0}, \quad (66f)$$

which are derived using first-order approximation according to Remark C.1. Note that we here consider only the case when (65a) and (65b) are active. Other cases (i.e., both are inactive, and either of (65a) and (65b) is active) can be simply treated.

It can be shown that, Eq. (66) is satisfied by taking $\{\mathbf{D_X}, \mathbf{D_F}, \mathbf{D_V}, \mathbf{D_{\bar{X}}}, \mathbf{D_{\bar{F}}}, \mathbf{D_{\bar{V}}}\} = \{-2\mathbf{X}, -\mathbf{F}, -\mathbf{V}, -2\mathbf{X}, -\mathbf{F}, -\mathbf{V}\}$ with $\mathbf{V} \neq \mathbf{0}$, and $\mathbf{F} \neq \mathbf{0}$ or $\mathbf{X} \neq \mathbf{0}$. This completes the proof. ∎

### C. CQ verification for problem (31)

For problem (31), we have

*Lemma C.3:* MFCQ holds for problem (31) at any feasible point $(\mathbf{X}, \mathbf{S}, \mathbf{Y})$.

*Proof:* Similarly as for problem (11), we can show that the linear independence of the equality constraint gradients of problem (31). So our main efforts are paid to show that, there exists $\{\mathbf{D_X}, \mathbf{D_S}, \mathbf{D_Y}\}$ such

$$\mathbf{Y D_S} + \mathbf{D_Y S} = \mathbf{0}, \quad (67a)$$
$$\mathbf{D_X} - \mathbf{D_Y} = \mathbf{0}, \quad (67b)$$
$$\mathbf{D_S}^T \mathbf{1} = \mathbf{0}, \quad (67c)$$
$$[\mathbf{D_S}]_{i,j} > 0, \forall (i,j) \in \mathcal{S}_0, \quad (67d)$$

where $\mathcal{S}_0$ is the set of zero entry indices of $\mathbf{S}$, $[\mathbf{D_S}]_{i,j}$ is the $(i,j)$-th entry of $\mathbf{D_S}$.

Let us check Eq. (67) with the point $(\mathbf{D_X}, \mathbf{D_Y}, \mathbf{D_S})$ given by

$$\mathbf{D_S} = \frac{1}{L}\mathbf{1}\mathbf{1}^T - \mathbf{S}, \mathbf{D_X} = \mathbf{D_Y} = \mathbf{Y}\left(\mathbf{I} - \frac{1}{L}\mathbf{1}\mathbf{1}^T\right). \quad (68)$$

where $\mathbf{1}$ is an all-one vector of dimension $L$. Obviously, Eqs. (67b) and (67d) are true. Furthermore, we have

$$\begin{aligned}\mathbf{D_S}^T \mathbf{1} &= (\frac{1}{L}\mathbf{1}\mathbf{1}^T - \mathbf{S})\mathbf{1} \\ &= \mathbf{1} - \mathbf{S}^T\mathbf{1} = 0\end{aligned} \quad (69)$$

$$\nabla h(\boldsymbol{w},t) = \begin{bmatrix} \frac{\boldsymbol{w}^T\mathbf{A}_1^T}{\|\mathbf{A}_1\boldsymbol{w}\|} - \frac{t_1\boldsymbol{w}^T\mathbf{B}_1^T}{\|\mathbf{B}_1\boldsymbol{w}\|} & -\|\mathbf{B}_1\boldsymbol{w}\| & \cdots & 0 \\ \vdots & \vdots & \vdots & \vdots \\ \frac{\boldsymbol{w}^T\mathbf{A}_K^T}{\|\mathbf{A}_K\boldsymbol{w}\|} - \frac{t_K\boldsymbol{w}^T\mathbf{B}_K^T}{\|\mathbf{B}_K\boldsymbol{w}\|} & 0 & \vdots & -\|\mathbf{B}_K\boldsymbol{w}\| \\ 2\boldsymbol{w}^T & \mathbf{0}^T & \mathbf{0}^T & \mathbf{0}^T \end{bmatrix}. \quad (64)$$

where the last equality follows from the feasibility. Substituting (68) into (67a), we obtain

$$\begin{aligned}
&\mathbf{Y}(\frac{1}{L}\mathbf{1}\mathbf{1}^T - \mathbf{S}) + \mathbf{D}_\mathbf{Y}\mathbf{S} \\
&= \frac{1}{L}\mathbf{Y}\mathbf{1}\mathbf{1}^T - \mathbf{Y}\mathbf{S} + \mathbf{D}_\mathbf{Y}\mathbf{S} \\
&= \frac{1}{L}\mathbf{Y}\mathbf{1}\mathbf{1}^T\mathbf{S} - \mathbf{Y}\mathbf{S} + \mathbf{D}_\mathbf{Y}\mathbf{S} \\
&= [\mathbf{D}_\mathbf{Y} - \mathbf{Y}(\mathbf{I} - \frac{1}{L}\mathbf{1}\mathbf{1}^T)]\mathbf{S} = \mathbf{0},
\end{aligned} \quad (70)$$

where the second equality is due to $\mathbf{S}^T\mathbf{1} = \mathbf{1}$. Therefore, MFCQ holds for problem (31) at any feasible point. ∎